\documentclass[preprint,12pt]{elsarticle}

\usepackage[margin=2.25cm]{geometry}
\usepackage{amssymb}
\usepackage[unicode=true]{hyperref}
\usepackage{amsmath}
\usepackage{braket}

\journal{Journal of Power Sources}

\begin{document}

\begin{frontmatter}



\title{A comprehensive first-principles investigation of A$_2$BH$_6$-type double perovskite hydrides (A = Li, Na, and K; B = Al, and Si) for high-capacity hydrogen storage}


\author[inst1,inst2]{R. Zosiamliana\corref{cor1}}
\author[inst2]{Lalhriat Zuala}
\author[inst2]{Shivraj Gurung}
\author[inst3]{R. Lalmalsawma}
\author[inst4]{A. Laref}
\author[inst5]{A. Yvaz}
\author[inst1]{D.P.Rai}
\affiliation[inst1]{organization={Department of Physics},
	addressline={Mizoram University}, 
	city={Aizawl},
	postcode={796004}, 
	country={India}}
\affiliation[inst2]{organization={Physical Sciences Research Center (PSRC), Department of Physics},
	addressline={Pachhunga University College}, 
	city={Aizawl},
	postcode={796001}, 
	country={India}}
\affiliation[inst3]{organization={Department of Computer Science},
	addressline={Government Champhai College}, 
	city={Champhai},
	postcode={796321}, 
	country={India}}
\affiliation[inst4]{organization={Department of Physics and Astronomy, College of Science},
	addressline={King Saud University}, 
	city={Riyadh},
	postcode={11451}, 
	country={Saudi Arabia}}
\affiliation[inst5]{organization={World-class research center "Advanced Digital Technologies", College of Science},
	addressline={State Marine Technical University}, 
	city={Saint Petersburg},
	postcode={190121}, 
	country={Russia}}
\cortext[cor1]{siama@pucollege.edu.in}

\begin{abstract}
Recent breakthroughs in vacancy-ordered double perovskite hydride materials have underscored their significant potential for integration into next-generation high-capacity hydrogen energy storage platforms. In the present study, extensive first-principles calculations leveraging both the GGA and hybrid-HSE06 exchange-correlation functionals are conducted to systematically explore the intrinsic properties of A$_2$BH$_6$ (A = Li, Na, and K; B = Al, and Si) complex hydrides, with a focus on rigorously assessing their viability as solid-state hydrogen storage media. Thermodynamic stability for each hydride is demonstrated and confirmed by negative formation energies, determined by both the GGA and HSE06 formalisms. Additionally, mechanical stability is validated through compliance with Born's stability criteria. Electronic properties analysis reveals a semi-conducting behavior in Si-based hydrides (A$_2$SiH$_6$), whereas Al-based (A$_2$AlH$_6$) analogues display metallic nature, regardless of the A-site atoms and functionals adopted. For the semi-conducting hydrides, we have observed higher optical absorption peak ($\alpha$ $>$ 10$^6$ cm$^{-1}$) in the UV regime indicating potential application in UV-optoelectronic devices. Furthermore, all studied compounds adhere to Debye's low-temperature specific heat behavior and converge to the classical Dulong-Petit limit at elevated temperatures, in accordance with fundamental thermodynamic principles. For hydrogen storage applications, both Al- and Si-based hydrides meet key benchmarks set by the U.S. Department of Energy (DOE), achieving gravimetric hydrogen capacities (C$_{wt}$\%) exceeding 5.5\% when A = Li or Na, and exhibiting volumetric hydrogen densities ($\rho_V$) greater than 40 g.H$_2$ L$^{-1}$. Among all studied hydrides, Li$_2$AlH$_6$ and Li$_2$SiH$_6$ emerge as the two most promising candidates due to their outstanding C$_{wt}$\% ($>$ 12\%), elevated $\rho_V$ ($>$ 140 g.H$_2$ L$^{-1}$)), and favorable hydrogen desorption temperature ranges (T$_D$ = 450 - 650 K).
\end{abstract}




\end{frontmatter}


\section{Introduction}
\label{Introduction}
The global shift toward sustainable and environmentally benign energy paradigms has catalyzed intensive research into advanced energy conversion and storage technologies \cite{Baxter2009,Sadeq2024,Hassan2024,Bertaglia2024,Nnabuife2025}. Among the diverse array of energy strategies, hydrogen-based energy systems, particularly solid-state hydrogen storage have garnered significant attention as promising alternatives, owing to their exceptionally high gravimetric energy density, widespread availability, and environmentally benign combustion profile \cite{Mahmood2021,Mustafa2024,Zelai2024,Ayyaz2025}. At present, the production of hydrogen fuel is primarily driven by processes reliant on fossil fuels and biomass. Conventional methods such as steam methane reforming (SMR), partial oxidation (POX), and autothermal reforming (ATR) of hydrocarbons, along with coal gasification techniques employing fixed and fluidized bed reactors, remain the most widely adopted. In parallel, biomass-based hydrogen generation through thermochemical and biological conversion routes offers a renewable yet still developing pathway. A more sustainable and environmentally benign approach involves water splitting facilitated by renewable energy inputs, particularly solar and wind. Among these, electrochemical water electrolysis, employing technologies such as alkaline electrolyzers (AE), proton exchange membrane (PEM) electrolyzers, and solid oxide electrolyzers (SOE), has demonstrated high energy conversion efficiencies (ranging from 62\% to 90\%) while generating near-zero greenhouse gas emissions. \cite{El-Shafie2019,El-Emam2019,Tremel2015} However, the practical implementation of hydrogen as a widespread energy vector is hindered by challenges in its storage and transportation.
\par From the extensive survey into recent literature, a range of two-dimensional (2D) materials, including graphene, hexagonal boron nitride (h-BN), MXenes, transition metal dichalcogenides (TMDs), borophene, and phosphorene, highlights as promising candidates for hydrogen storage applications. \cite{Boateng2023,Chettri2021b,Kumar2021,Bora2024,Ledwaba2023,Garara2019} This potential stems from their high surface area, tunable electronic properties, and capacity for reversible hydrogen adsorption. Nevertheless, a significant challenge limiting their practical deployment is the structural degradation at high hydrogen coverage, which compromises their mechanical stability and adsorption reversibility, thereby restricting their viability for hydrogen storage systems. Therefore, the three-dimensional (3D) solid-state hydrogen storage materials, particularly complex metal hydrides, intermetallics, and nanostructured systems, are being extensively investigated for their potential to meet these stringent performance criteria, offering safer, more compact, and energy-efficient means of storing hydrogen \cite{Sakintuna2007,Kalibek2024,Nemukula2025}. Despite substantial advancements in these domains, formidable scientific and engineering challenges remain. In the context of hydrogen storage, the pursuit of materials that combine high gravimetric and volumetric capacities with favorable thermodynamics and rapid kinetics continues to be a critical research frontier. Addressing these challenges is essential for advancing the commercial viability of hydrogen-based energy solutions, thereby contributing to a more sustainable and resilient global energy infrastructure. Among the various material candidates, vacancy-ordered double perovskite hydrides (VODPH) (A$_2$BH$_6$) have attracted growing attention due to their intrinsic compositional versatility, tunable structural and electronic properties, and multifunctional applicability. These materials demonstrate remarkable potential into solid-state hydrogen storage, positioning them as strategic enablers for next-generation energy solutions \cite{Hakami2024,Alkhaldi2025}.
\par Despite their promising attributes, a comprehensive theoretical understanding of the high-capacity hydrogen storage application of the selected VODPH materials remain limited. This study aims to bridge this gap by employing a Density Functional Theory (DFT)-based investigation into the solid-state hydrogen storage of A$_2$BH$_6$ (A = Li, Na, and K; B = Al, and Si). The insights gained from this work could contribute to the development of next-generation materials for sustainable energy applications.
\section{Computational Details}
\label{Computational Details}
All simulations were performed within the DFT framework using an advanced software package, the MedeA VASP 5.4, which rely on the Projector Augmented Wave (PAW) method for efficient core-electron interactions in plane-wave-based electronic structure calculations \cite{Hafner2008a,Kresse1996e}. The PAW\_PBE Li\_sv 10Sep2004, PAW\_PBE Na\_pv 19Sep2006, PAW\_PBE K\_sv 06Sep2000, PAW\_PBE Al 04Jan2001, PAW\_PBE Si 05Jan2001, and PAW\_PBE H 15Jun2001 pseudopotentials were sampled for the Li-, Na-, K-, Al-, Si-, and H-atoms, respectively. Exchange-correlation interactions were initially modeled with the Perdew-Burke-Ernzerhof (PBE) functional within the generalized gradient approximation (GGA), as a baseline for comparison \cite{Perdew1996l}. To mitigate delocalization errors and improve property predictions, the range-separated hybrid screened-Coulomb (SC) functional HSE06 was employed \cite{Heyd2003c}. By incorporating exact Hartree-Fock exchange in the short range while maintaining a screened Coulomb potential at long range, HSE06 significantly enhances the accuracy of band gap estimations, charge density distributions, and the electronic structure's related properties.
\par The mathematical expressions for the hybrid-HSE06 functional employed is:
\begin{equation}
		E^{HSE06}_{XC}=\frac{1}{4}E^{SR, HF}_{X} (\omega)+\frac{3}{4}E^{SR, PBE}_{X} (\omega)+E^{LR, PBE}_{X} (\omega)+E^{PBE}_{C} 
\end{equation} 
\par Here, $\omega$ = 0.2 denotes the screening parameter. The term E$^{SR, HF}_{X}$ signifies the short-range Hartree-Fock exact exchange functional, while E$^{SR, PBE}_{X}$, and E$^{LR, PBE}_{X}$ correspond to the short-range and long-range PBE exchange functionals, respectively. Additionally, E$^{PBE}_{C}$ represents the complete correlation energy.
\par In this study, we delineated the bulk fundamental properties of A$_2$BH$_6$ VODPH (A = Li, Na, and K; B = Al, and Si), encompassing structural stability, electronics, optical, elastic, thermodynamics, and their energy application towards solid-state hydrogen storage using the GGA and the hybrid-HSE06 method, as implemented in MedeA VASP package. During bulk structural optimization, full-cell relaxation was performed, allowing atomic positions, unit cell shape, and volume to adjust. Energy convergence was ensured with a stringent criterion of 10$^{-6}$ eV, maintaining atomic forces below 0.02 eV/{\AA}. A plane-wave energy cutoff of 500 eV was employed for electronic wave function expansion. Optimization utilized analytical gradients with respect to atomic and cell parameters within a quasi-Newtonian framework, incorporating the Broyden-Fletcher-Goldfarb-Shanno (BFGS) updating method \cite{Head1985a,Nawi2006a}. Reciprocal space integration employed a Monkhorst-Pack k-mesh of 10 x 10 x 10 \cite{Monkhorst1976k,Blochl1994d}. The property calculations followed similar self-consistent field (SCF) criteria with an enhanced k-mesh of 12 × 12 × 12.
\section{Results and Discussions}
\subsection{Structural Stability and Electronic Properties}
\label{Structural Properties and Stability}
\begin{figure}[hbt!]
	\centering
	\includegraphics[height=4cm]{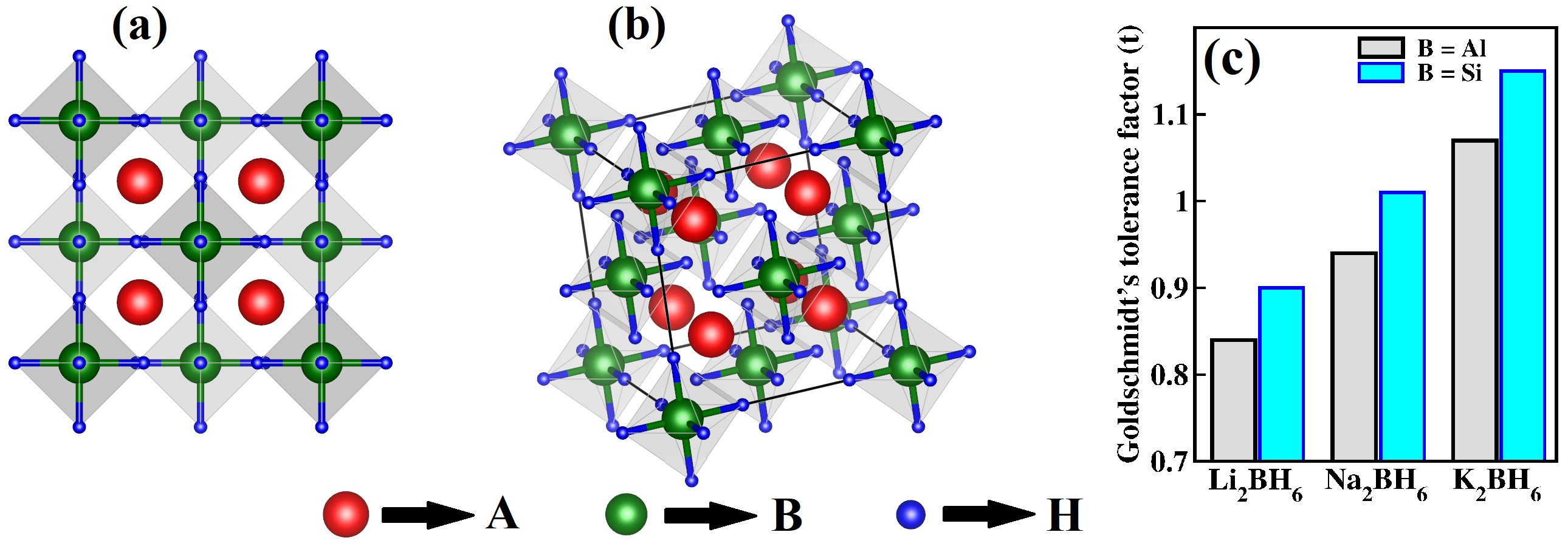}
	\caption{(a) and (b) represent the 3D- and 2D-crystallographic structures of A$_2$BH$_6$ VODPH (A = Li, Na, and K; B = Al, and Si), plotted using VESTA program. (c) The Goldschmidt's tolerance factor (t).}
	\label{figure1}
\end{figure}
Within the framework of the GGA based DFT methodology, the optimized structures of the VODPH we considered in this work exhibit cubic phase crystallization, corresponding to $Fm$-3$m$ (No. 225) space group, as illustrated in figure \ref{figure1} (a) and (b). The structural configuration of the studied A$_2$BH$_6$ hydrides follows a well-defined atomic arrangement: the A-site atoms occupy the 8c (0.25, 0.25, 0.25), B-site atoms reside at 4a (0.00, 0.00, 0.00), and hydrogen atoms are positioned at 24e (xx, 0.50, 0.50) Wyckoff positions and fractional coordinates \cite{Talebi2023,Alburaih2024,Hayat2023}. The structural parameters, including GGA optimized lattice constants, the formation energy, and electronic band gaps calculated from the GGA and hybrid-HSE06 formalisms, are systematically presented in table \ref{table1}. Evidently, a distinct trend is observed in the variation of lattice parameters with atomic substitution. Specifically, as the A-site cation moves down the group (i.e., from Li $\rightarrow$ K), the lattice constant exhibits a systematic increase, which can be attributed to the progressive enlargement of ionic radii down the group. In contrast, the Al-based hydrides demonstrate larger lattice constants compared to their Si-based counterparts. This disparity arises from the substitution of the B-site cation across the period (i.e., Al $\rightarrow$ Si), wherein the lattice constants decrease due to the reduced atomic radius and the strengthening of bonding interactions along the period.
\begin{table}[hbt!]
	\small
	\caption{\ The optimized lattice parameter (a) (in {\AA}), the formation energy (E$^f$) (in eV), and the electronic band gaps (E$_g$) (in eV) as obtained from both the GGA and HSE06 methods.}
	\label{table1}\renewcommand{\arraystretch}{1.55}
	\begin{tabular*}{\textwidth}{@{\extracolsep{\fill}}l|lllll|lllll}
	\hline
	&&& B = Al &&&&& B = Si &&\\
	A = & a & E$^f_{GGA}$ & E$^f_{HSE06}$ & E$^{GGA}_g$ & E$^{HSE06}_g$ & a & E$^f_{GGA}$ & E$^f_{HSE06}$ & E$^{GGA}_g$ & E$^{HSE06}_g$\\ \hline  
	Li &6.55&-0.66&-0.71&--- ---&--- ---&6.53&-0.77&-0.87&1.03&2.08\\ 
	Na &7.16&-0.74&-0.75&--- ---&--- ---&7.11&-0.88&-0.95&1.29&2.37\\
	K  &7.93&-0.91&-0.79&--- ---&--- ---&7.77&-1.09&-1.04&2.23&3.38\\
	\hline
	\end{tabular*}
\end{table}
\par To rigorously assess the thermodynamic stability and validate the crystallization of the investigated perovskite hydrides, the formation energy (E$^f$) and Goldschmidt's tolerance factor (t) are calculated using equation \ref{equation2} \cite{Renthlei2023h,Zosiamliana2025a}:
\begin{equation}
	\label{equation2}
	\begin{split}
		E^f=\frac{\big[E_T-\big(2\times E_A+E_B+6\times E_H\big)\big]}{n} \\
		\tau=\frac{r_A+r_H}{\sqrt{2}(r_B+r_H)}	
	\end{split}
\end{equation}
Here, E$_T$ denotes the total ground-state energy, while E$_A$, E$_B$, and E$_H$ correspond to the atomic energies of the A-site, B-site, and hydrogen atoms, respectively. Also, $r_A$, $r_B$, and $r_H$ represent the ionic radii of A- and B-atoms, and the hydrogen ionic radius, respectively. 
\par The formation energy (E$^f$) calculated using both the GGA and HSE06 approaches, as summarized in table \ref{table1}, consistently yields negative values for all examined perovskite compounds. This strongly suggests their thermodynamic stability and an energetically favorable (exothermic) formation process. A comparative analysis reveals that the Si-based VODPH compounds exhibit superior stability and formability relative to their Al-based counterparts, which can be ascribed to their relatively higher (more negative) formation energy. Furthermore, for both Al- and Si-based hydrides, the substitution of A-site cations down the group leads to an enhancement in thermodynamic stability, reflecting the influence of increasing ionic radii and lattice expansion. However, an intriguing trend emerges upon employing the hybrid-HSE06 functional: while it further stabilizes the structures for A = Li and Na, a reduction in stability is observed when A = K. This suggests that the effect of hybrid functional corrections on formation energy is composition-dependent, potentially influenced by variations in electronic interactions and bonding characteristics across different A-site cations. As depicted in figure \ref{figure1}, the Goldschmidt's tolerance factor (t), calculated using equation \ref{equation2}, provides additional insight into the structural viability of the examined materials. The analysis indicates that all investigated compounds adopt a perovskite structure in their pristine state, with the exception of K$_2$AlH$_6$ and K$_2$SiH$_6$, which exhibit tolerance factors outside the stability range (t $<$ 0.7 or t $>$ 1), thereby precluding a perovskite configuration. Notably, among the studied hydrides, Na$_2$SiH$_6$ is found to crystallize in an ideal cubic perovskite structure, characterized by a tolerance factor of precisely t = 1, indicating optimal geometric compatibility within the perovskite framework.
\begin{figure}[hbt!]
	\centering
	\includegraphics[height=8.25cm]{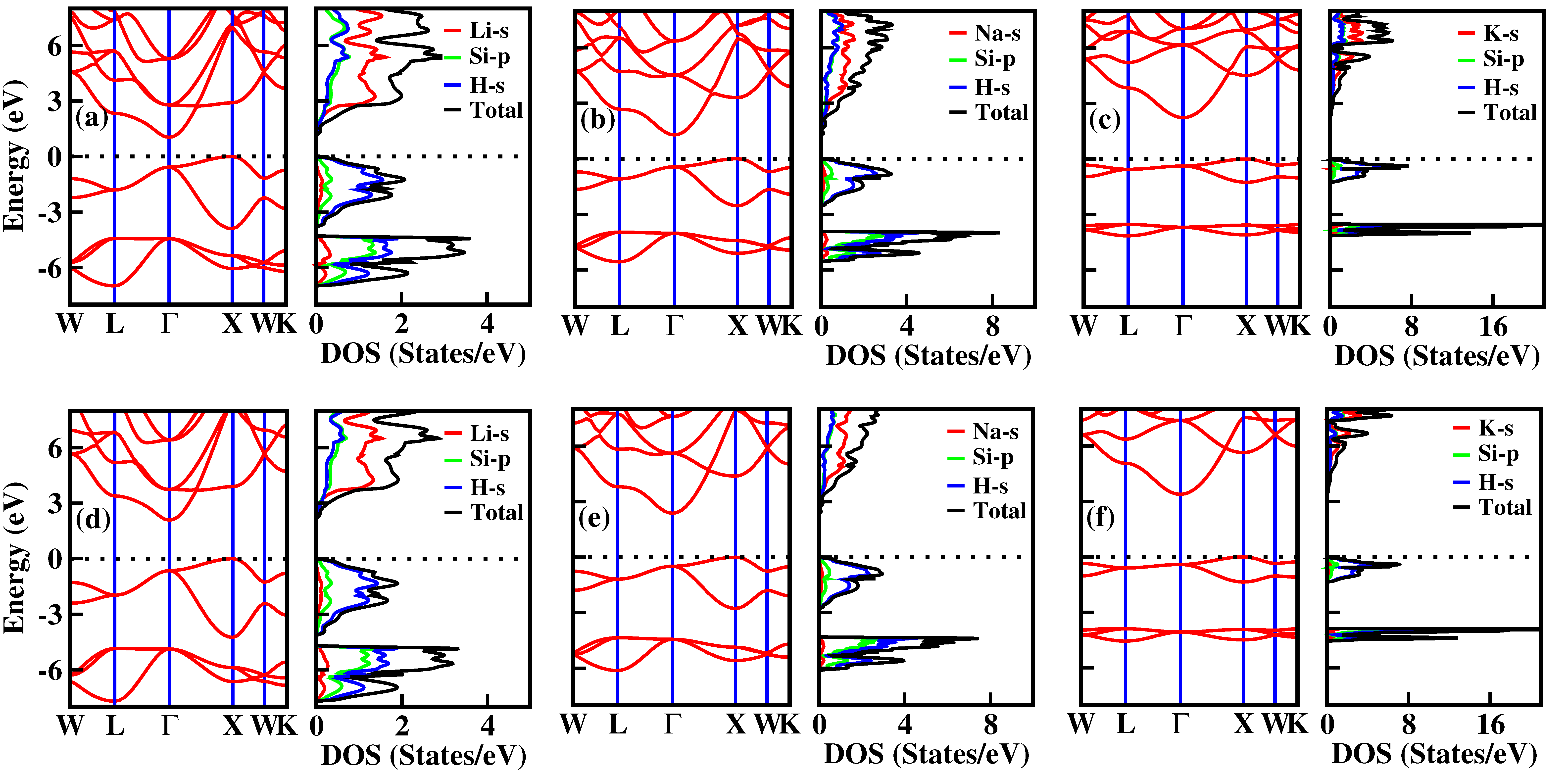}
	\caption{Electronic band structures and density of states (DOS) of A$_2$SiH$_6$ VODPH (A = Li, Na, and K) computed using (a)-(c) GGA and (d)-(e) HSE06 functionals. The green dotted line at 0 eV energy represents the Fermi energy (E$_F$).}
	\label{figure2}
\end{figure}
\par Based on the GGA and hybrid-HSE06 DFT methodologies, the electronic properties of the investigated hydride materials; specifically, the band structures, density of states (DOS), and charge transfer (Q$_T$), are systematically computed \cite{Mulliken1955b}. These properties are essential for elucidating the underlying atomic-level interactions, providing critical insights into the electronic structure, bonding characteristics, and charge redistribution mechanisms within the studied compounds.
\par The electronic band gap (E$_g$) estimated using both the GGA and HSE06 functionals at high-symmetry points, as summarized in table \ref{table1}, along with the electronic band structures and density of states (DOS) depicted in figures S1 and \ref{figure2}, reveal distinct electronic properties for the investigated materials. Specifically, the Al-based hydrides exhibit a metallic nature, indicating a finite density of states at the Fermi level (E$_F$). In contrast, the Si-based counterparts display an indirect X $\rightarrow$ $\Gamma$ semi-conducting behavior, suggesting a band gap transition between the conduction band minimum (CBM) and valence band maximum (VBM) occurring at different k-points in the Brillouin zone. For the Al-based hydrides, as observed in figure S1, the metallic nature arises due to the crossing of E$_F$ by the VBM at the X- and $\Gamma$-symmetry points. This behavior is primarily driven by the significant DOS contributions from the hybridized Al-3p and H-1s orbitals. Conversely, for the semi-conducting Si-based compounds the VBM and CBM near the E$_F$ are mostly contributed by the complex hybridized states of the A-s, Si-3p, and H-1s states. The disparity in electronic properties between the Al- and Si-based hydrides can be fundamentally understood through the computed charge transfer (Q$_T$) values presented in table S1. In Al-based hydrides, the electronic structure is characterized by a significant hybridization between Al-3p and H-1s states. This strong orbital interaction induces substantial band broadening, leading to an overlap between the valence and conduction bands at the E$_F$. As a result, these systems exhibit metallic behavior. Conversely, in Si-based hydrides, the Si-3p orbitals hybridize with H-1s states in a predominantly covalent manner, which enhances electronic localization and reinforces the energetic separation between the CBM and VBM, thereby imparting a semi-conducting nature to the material. Mulliken population analysis further reveals a fundamental contrast in charge transfer mechanisms between these two hydride systems. In Al-based hydrides, the alkali metal atoms (Li, Na, K) exhibit charge depletion, with Q$_T$ (A) = 0.33 to 0.66 $\vert$e$\vert$, while Al-atoms similarly experience a net positive charge transfer, with Q$_T$ (Al) = 0.68 and 0.82 $\vert$e$\vert$. This charge redistribution is counter-balanced by the hydrogen atoms, which accumulate excess electrons (Q$_T$ (H) = -0.23 to -0.36 $\vert$e$\vert$), a characteristic commonly observed in perovskite-derived materials. In contrast, Si-based hydrides exhibit a more complex charge transfer behavior due to the higher electronegativity of Si relative to Al. The alkali metal atoms in these systems undergo even greater charge depletion (Q$_T$ (A) = 0.56 to 0.78 $\vert$e$\vert$). Moreover, Si, having an electronegativity closer to that of hydrogen, exhibits an unconventional charge redistribution. Specifically, for A = Li and Na compounds, Si-atoms gain electrons (Q$_T$ (Si) = -0.08 to -0.25 $\vert$e$\vert$), while in K$_2$SiH$_6$ hydride, Si-atoms experience a slight charge loss (Q$_T$ (Si) = 0.07 to 0.15 $\vert$e$\vert$), indicating that Si does not function as a strongly electropositive cation in these Si-based hydrides. The charge imbalance in the systems are ultimately compensated by hydrogen atoms (Q$_T$ (H) = -0.15 to -0.28 $\vert$e$\vert$). This analysis suggests that the Al-H bonding exhibits a degree of electronic delocalization, enabling partial metallic conductivity for A$_2$AlH$_6$ hydrides, while the Si-H interaction remains highly covalent and localized, which significantly restricts free-electron mobility and stabilizes a band gap in A$_2$SiH$_6$, reinforcing its semi-conducting nature. 
\subsection{Optical properties}
\label{Optical Properties}
\begin{figure}[hbt!]
	\centering
	\includegraphics[height=8.25cm]{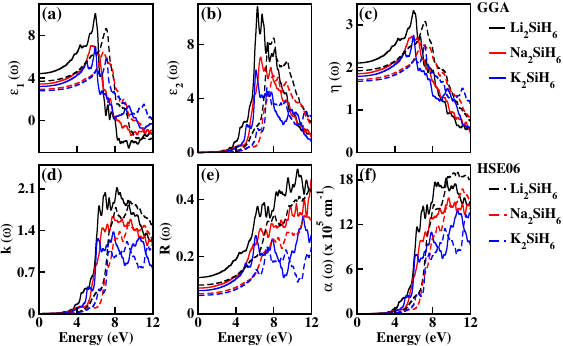}
	\caption{Calculated optical properties of A$_2$SiH$_6$ VODPH (A = Li, Na, and K) computed using GGA and HSE06 functionals. (a) Real part of dielectric constant ($\epsilon_1$), (b) Imaginary part of dielectric constant ($\epsilon_2$), (c) Refractive index ($\eta$), (d) Extinction coefficient (k), (e) Reflectivity (R), and (f) Absorption coefficient ($\alpha$)}
	\label{optical}
\end{figure} 
To determine the interplay between the incoming electromagnetic radiation and the studied hydride materials, we have presented the optical properties by calculating the complex dielectric constants, refractive index, absorption coefficient, and other parameters as a function of photon energy (eV), using the GGA and hybrid-HSE06 functionals. As mentioned in the above section (No. \ref{Structural Properties and Stability}), the semi-conducting nature of the investigated hydrides are demonstrated when the B-site atom is silicon (Si). Therefore, the optical properties and thereby their potential opto-electronic device applications are only explored for the semi-conducting hydrides. The complex dielectric function consisting of the real part ($\epsilon_1$) and the imaginary part ($\epsilon_2$) are determined as: \cite{Ambrosch2006}  
\begin{equation}
	\epsilon = \epsilon_1 + i\epsilon_2 
\end{equation}
Here, the real part of the dielectric function, $\epsilon_1$, is derived from the imaginary part, $\epsilon_2$, via the Kramers-Kronig's transformation, based on the classical oscillator model that describes the vibrational and electronic response of atoms, notify as:
\begin{equation}
	\epsilon_1(\omega) = 1 + \frac{2}{\pi}\int_{0}^{\infty}\frac{\epsilon_2 (\omega') \omega' d\omega'}{\omega'^2 - \omega^2}
\end{equation}
\begin{equation}
	\epsilon_2(\omega) = \frac{\hbar^2 e^2}{\pi m^2 \omega^2}\sum_{nn'}\int_{k} d^3k\bigg \vert \braket{\vec{k}_n \vert \vec{p} \vert \vec{k'}_{n}}\bigg \vert ^2 \times \bigg[1-f(\vec{k}_n)\bigg]\delta(E_{\vec{k}_n} - E_{\vec{k'}_{n}} - \hbar \omega)
\end{equation} 
Where, $\vec{p}$ is momentum operator, $\ket{\vec{k}_n}$ is eigenfunction of eigenvalue E$_{\vec{k}_n}$ and $f(\vec{k}_n)$ is Fermi distribution function.
\par The employed formulae for other optical parameters including; refractive index ($\eta$), extinction coefficient (k), optical reflectivity (R), and absorption coefficient ($\alpha$) are:
\begin{equation}
	\begin{split}
		\eta(\omega) = \sqrt{\frac{(\epsilon_1^2 + \epsilon_2^2)^\frac{1}{2} + \epsilon_1}{2}}\\
		k(\omega) = \sqrt{\frac{(\epsilon_1^2 + \epsilon_2^2)^\frac{1}{2} - \epsilon_1}{2}}\\
		R(\omega) = \frac{(1-\eta)^2+k}{(1+\eta)^2+k}\\
		\alpha(\omega) = 2\frac{\omega}{c}k(\omega)
	\end{split}
\end{equation}
\par In figure \ref{optical} (a), the real part of the dielectric constant, $\epsilon_1$, is plotted as a function of photon energy, which is intrinsically inter-linked to the material's refractive index ($\eta$), as illustrated in figure \ref{optical} (c). The $\epsilon_1$ spectrum reflects the material's polarization response towards an external electromagnetic radiation, thus providing key insights into the optical anisotropy and dielectric screening behavior. At the static limit (zero photon energy), the calculated static real dielectric constants, $\epsilon_1(0)$, are found to be 4.43, 3.43, and 3.20 arb. units for Li$_2$SiH$_6$, Na$_2$SiH$_6$, and K$_2$SiH$_6$, respectively, within the framework of standard GGA. In comparison, the hybrid-HSE06 results in slightly lower values of 3.70, 2.94, and 2.81 arb. units for the corresponding similar compounds, indicating the impact of improved exchange-correlation treatment on electronic screening properties. Correspondingly, the calculated static refractive indices ($\eta(0)$) are 2.10, 1.85, and 1.79 for the GGA calculations, and 1.92, 1.71, and 1.67 under the HSE06 approach, respectively. These values reveal that the studied materials are translucent or opaque in nature. The most significant peaks for both $\epsilon_1$ and $\eta$ appear within 5.5 to 7.5 eV in the UV spectral region, and notably, these peak positions exhibit a subtle blue shift as A-side cation goes from Li $\rightarrow$ K. This shift suggest an enhanced optical transition energies, attributes to varying electronic structure profile. The negative $\epsilon_1$ values existed at higher energy region ($\omega$ $>$ 8.0 eV) reveals the loss of light transit, essentially caused by plasmonic oscillations. \cite{WANG1996} This subsequently let $\eta$ to reduced below 1 at higher photon energy region.   
\par The $\epsilon_2$ which is inter-connected to the extinction coefficient (k) and thereby the optical absorption spectrum ($\alpha$) are illustrated in figures \ref{optical} (b), (d), and (f), respectively. These optical parameters provide direct insights into the interband electronic transitions between the valence and conduction bands. The calculated onset energies, corresponding to the threshold of interband transitions, for A = Li, Na, and K within the GGA framework are 3.36, 3.94, and 4.53 eV, respectively. These values exhibit a noticeable blue shifting upon computing with hybrid-HSE06 functional, with the corresponding threshold energies shifting to 4.26 eV, 5.16 eV, and 5.67 eV, respectively. These optical thresholds correspond well to the direct optical band gaps and are consistent with the calculated electronic structures. Specifically, for Li$_2$SiH$_6$, the transition occurs from the valence band maximum to the second conduction band along the $\Gamma$-symmetry point. In contrast, for Na$_2$SiH$_6$ and K$_2$SiH$_6$, the lowest-energy direct transitions relevant to the optical band gaps take place at the L-symmetry point, involving the first bands of both the valence and conduction regions. Despite the quantitative discrepancies in the threshold energies between the GGA and hybrid-HSE06 methods, the nature of the transitions, both in terms of symmetry points and involved band indices, remains consistent across the two approaches, confirming the robustness of the band structure and optical response analysis. Given that the most prominent features of the $\epsilon_2$ and k are located in the UV-region, the investigated semiconducting perovskite hydrides A$_2$SiH$_6$ (A = Li, Na, and K) exhibit strong absorption capability for UV radiation, indicating their potential effectiveness in UV-active optoelectronic applications.
\par In this section, particular emphasis is placed on evaluating key optical parameters namely, the optical reflectivity (R) and $\alpha$, which are crucial for assessing material performance in optoelectronic applications. As illustrated in figure \ref{optical} (e), the reflectivity spectra as a function of photon energy, reveal a systematic trend for the studied semiconducting Si-based hydrides. Specifically, as the A-site cation varies from Li $\rightarrow$ K, a discernible reduction in overall reflectivity is observed. This decline is attributed to the increasing ionic radius and corresponding changes in electronic structure, which alter the material's interaction with incident radiation. Also, upon hybrid-HSE06 calculation each compound's reflectivity consistently minimizes compared to the GGA estimated results, indicating more accurate depiction of electronic excitations using HSE06 functional. The static reflectivity (R (0)) values range between 0.08 to 0.13 using GGA framework, while these values drop between 0.06 to 0.10 under hybrid-HSE06 method. In the UV spectral region (6.0 $\le$ $\omega$ $\le$ 12.0 eV), all compounds exhibit significant reflectance peaks, reaching up to $\sim$50 \% from $\sim$20 \%. This underscores their potential for use in UV-reflective or filtering optical devices. Complementary insights for optoelectronic applications are provided by the $\alpha$ spectra, as presented in figure \ref{optical} (f), which follow a consistent trends with the $\epsilon_2$ and the k. The $\alpha$ spectra display a rapid increase with photon energy, indicative of intense interband electronic transitions, particularly in the UV regime. A blue shift in the energy threshold of $\alpha$ is observed as the A-site atom transitions from Li $\rightarrow$ K, pointing to a widening of the optical band gap due to modifications in the electronic band structure. The threshold absorption energies, spanning from 3.0 to 6.0 eV (across both GGA and HSE06 frameworks), confirm that these hydrides are capable of absorbing photons from the upper visible to UV regions. Above 5.0 eV, $\alpha$ increases sharply, but the magnitude of absorption begins to attenuate with heavier A-site atoms, suggesting a correlation between ionic size and optical transition probabilities. The most dominant absorption features occur within the 8.0 to 12.0 eV range-well into the deep UV spectrum, with $\alpha$ exceeding 10$^6$ cm$^{-1}$. These such high absorption intensities highlight the materials' potential suitability for incorporation in UV-active optoelectronic devices, such as UV photo-detectors, high-frequency photonic components, etc. Overall, the hybrid-HSE06 functional provides a more physically accurate representation of the optical behavior by correcting the band gap underestimation inherent to GGA, however yield less intense optical response. This trade-off emphasizes the importance of functional selection in theoretical predictions for optical material design and engineering.
\subsection{Elastic properties}
\label{Elastic Properties}
The evaluation of mechanical properties is crucial for determining mechanical stability of compounds. Given that the examined materials crystallize in the cubic phase, their elastic behavior is characterized by three independent elastic constants: C$_{11}$, C$_{44}$, and C$_{12}$. The necessary conditions for mechanical stability in cubic systems, known as the Born criteria are \cite{Mouhat2014j,Born1940i}:
\begin{equation}
	\begin{split}
		C_{11}>1, C_{44}>1, C_{11}-C_{12}>1\\
		C_{11}+2C_{12}>1, C_{12}<B<C_{11}		
	\end{split}
\end{equation}
\begin{table}[hbt!]
	\small
	\caption{\ Calculated elastic constants (C$_{ij}$) and Cauchy's pressure (C$_{12}$-C$_{44}$) in GPa unit.}
	\label{table2}\renewcommand{\arraystretch}{1.55}
	\begin{tabular*}{\textwidth}{@{\extracolsep{\fill}}l|llll|llll}
		\hline
		&   & B = Al &    &         && B = Si      &       &\\ \hline    
		A = &  C$_{11}$ & C$_{44}$ & C$_{12}$ & C$_{12}$-C$_{44}$ & C$_{11}$ & C$_{44}$ & C$_{12}$ & C$_{12}$-C$_{44}$\\
		\hline
		Li & 31.83 & 39.30 & 25.69 & -13.61 & 46.83 & 37.92 & 18.71 & -19.21 \\
		Na & 27.57 & 29.29 & 18.42 & -10.87 & 41.33 & 32.85 & 11.31 & -21.54 \\
		K  & 30.28 & 22.57 & 12.49 & -10.08 & 40.06 & 26.40 & 7.32  & -18.78 \\
		\hline
	\end{tabular*}
\end{table}
\par The computed elastic constants (C$_{ij}$) and bulk modulus align with established stability criteria, confirming the mechanical stability of the investigated materials. As presented in table \ref{table2}, the elastic constants: C$_{11}$, C$_{44}$, and C$_{12}$, represent longitudinal, shear, and transverse (coupling) elastic moduli, respectively. These parameters define resistance to uniaxial strain (C$_{11}$), shear deformation (C$_{44}$), and lateral deformation effects governed by Poisson’s ratio (C$_{12}$). A comparative analysis with prior studies reveals distinct mechanical responses between the studied VODPH materials compared to conventional perovskite hydrides (ABH$_3$) examined by Zosiamliana \textit{et al.,} \cite{Zosiamliana2025a} and ordered double perovskite hydrides KNaX$2$H$6$ (X = Mg, Ca) investigated by Rehman \textit{et al} \cite{Rehman2024}. In the reference systems, a pronounced C$_{11}$ $>>$ C$_{44}$ trend indicates higher resistance to axial compression than shear deformation, with bulk modulus (B) exceeding shear modulus (G). However, the VODPH materials exhibit anomalous elastic trends. For Li$_2$AlH$_6$ and Na$_2$AlH$_6$, an unconventional C$_{11}$ $<$ C$_{44}$ relationship is observed, despite B $>$ G (see table S2), indicating more resistant to volume change than to shear deformation. This behavior likely arises from asymmetric bonding within the vacancy-ordered lattice, inducing directional anisotropy. Conversely, Na$_2$SiH$_6$ and K$_2$SiH$_6$ display C$_{11}$ $>>$ C$_{44}$ yet B $<$ G, signifying dominant shear resistance over volumetric rigidity, likely due to B-site cation radius variations affecting octahedral flexibility. Meanwhile, K$_2$AlH$_6$ and Li$_2$SiH$_6$ adhere to conventional elastic trends, aligning with other perovskite type hydrides, indicating balanced stress distribution within their octahedral framework. The mechanical anomalies in the studied A$_2$BH$_6$ VODPH structures, compared to A$_2$BB'H$_6$ ordered double perovskites, stem from atomic-scale structural differences. In A$_2$BH$_6$, a single B-site cation establishes a cubic BH$_6$ octahedral network, but vacancy-induced disruptions impart lattice anisotropy, weakening structural connectivity and altering stress propagation. Additionally, small A-site and B-site cations reduce steric effects, enhancing lattice flexibility while modifying the force distribution network. The absence of a fully interconnected octahedral framework fundamentally reshapes mechanical responses, driving elastic anisotropy and stability variations across compositions. Meanwhile, the Young's modulus (Y) presented in table S2 indicates Si-based hydrides exhibiting higher Y values than their Al-based counterparts, signifying superior mechanical stiffness and greater resistance to elastic deformation. Additionally, as B-site element moves down the group, Si-based hydrides demonstrate a consistent decline in Y values, whereas Al-based hydrides exhibit non-monotonic fluctuations. This contrasting trend underscores the complex interplay of atomic size, and bonding characteristics in governing the mechanical properties of these hydrides.
\par The failure mechanisms of the materials, characterized by their brittleness and ductility, along with key mechanical parameters including internal deformation ($\zeta$), elastic anisotropy (A$_{an}$), machinability factor ($\mu_m$), and melting temperature (T$_m$) are illustrated in figure S2 and summarized in table S3 (refer to equations S1-S4) \cite{Kleinman1962h,Chung1968f,Sun2005c,Ahmed2023c}. Based on the Pugh's ratio and Poisson's ratio plotted in figure S2, all the examined hydride materials exhibit brittle behavior, with Si-based hydrides demonstrating the higher brittleness. Furthermore, as the B-site element descends down the group, the brittleness of the corresponding hydrides intensifies. This brittle nature is further corroborated by the negative Cauchy's pressure values presented in table \ref{table2}. The computed $\zeta$ reveals that in Al-based hydrides, bond bending predominantly contributes to internal deformation ($\zeta \rightarrow 1$), whereas in Si-based hydrides, bond stretching plays a more significant role ($\zeta \rightarrow 0$). Additionally, the calculated A$_{an}$ indicates that Li$_2$AlH$_6$ and Na$_2$AlH$_6$ exhibit pronounced elastic anisotropy compared to other compounds, consistent with the anomalous elastic trends discussed earlier. Furthermore, the $\mu_m$ and T$_m$ values suggest that all the studied materials possess moderate machinability and melting points. The major concern for calculating the mechanical properties is to determine the Debye's temperature ($\Theta_D$) and the sound velocities, presented in table S3 (refer to equations S5-S8) \cite{Cahill1989i,Tani2010i,Chen2011l}. The investigated compounds exhibit significantly higher $\Theta_D$ compared to various crystal structures, including Pb-based oxide perovskites PbXO$_3$ (X = Ti, Zr, and Hf), which typically range between 400-500 K \cite{Renthlei2023h}. However, their $\Theta_D$ values are comparable to those of single hydride perovskites ABH$_6$. A higher $\Theta_D$ signifies stronger interatomic interactions and bonding strength. In all cases, V$_l$ $>$ V$_t$ and, as the B-site atoms descend down the group, sound velocity exhibits a continuous decline for Si-based hydrides, whereas it fluctuates for their Al-based counterparts. Additionally, the average sound velocity (V$_{av}$) of Si-based hydrides surpasses that of the corresponding Al-based compounds. Overall, the computed mechanical properties strongly support the potential of these materials for the solid-state hydrogen storage and thermoelectric applications.
\subsection{Thermodynamics properties}
\label{Thermodynamics Properties}
\begin{figure}[hbt!]
	\centering
	\includegraphics[height=8cm]{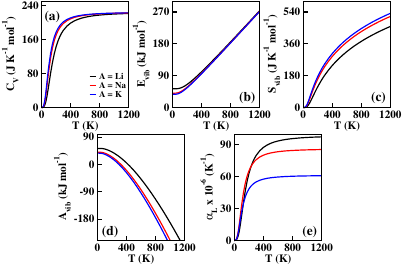}
	\caption{Variations of (a) constant-volume heat capacity C$_V$, (b) change in vibrational internal energy E$_{vib}$, (c) vibrational entropy S$_{vib}$, (d) vibrational Helmholtz free energy A$_{vib}$, and (e) linear thermal expansion coefficient $\alpha_L$ with respect to temperature T (K) for A$_2$AlH$_6$ (A = Li, Na, and K) hydrides.}
	\label{figure3}
\end{figure}
\begin{figure}[hbt!]
	\centering
	\includegraphics[height=8cm]{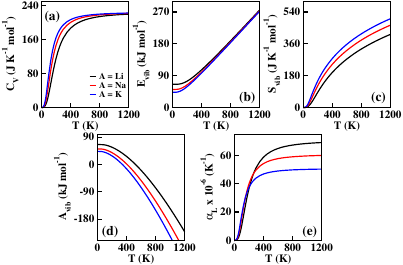}
	\caption{Variations of (a) constant-volume heat capacity C$_V$, (b) change in vibrational internal energy E$_{vib}$, (c) vibrational entropy S$_{vib}$, (d) vibrational Helmholtz free energy A$_{vib}$, and (e) linear thermal expansion coefficient $\alpha_L$ with respect to temperature T (K) for A$_2$SiH$_6$ (A = Li, Na, and K) hydrides.}
	\label{figure4}
\end{figure}
\par To thoroughly assess the thermodynamic characteristics that elucidate the energetic state of the studied systems, we present a detailed analysis of their vibrational properties. Specifically, we report the temperature-dependent behavior of the constant-volume specific heat (C$_V$), vibrational internal energy (E$_{vib}$), vibrational entropy (S$_{vib}$), vibrational Helmholtz free energy (A$_{vib}$), and the linear thermal expansion coefficient ($\alpha_L$) for all investigated hydrides. These properties were computed within the framework of the quasi-harmonic Debye model, which accounts for anharmonic lattice vibrations and their influence on thermodynamic responses (see equation S9-S12) \cite{Chen2001c,Blanco2004g}.
\par The extent of intrinsic disorder within a system, often quantified by the E$_{vib}$, S$_{vib}$, and A$_{vib}$, serves as a key indicator of thermodynamic randomness. As illustrated in figures \ref{figure3} and \ref{figure4}, for both Al- and Si-based hydrides, a progressive substitution of the A-site atom from Li $\rightarrow$ K leads to a decrease in E$_{vib}$ and A$_{vib}$, accompanied by an increase in S$_{vib}$. This trend implies a notable reduction in the thermodynamic potential or the fraction of thermal energy available to perform useful work. Such behavior arises from the enhanced dispersal of thermal energy within the lattice, primarily driven by the increasing atomic radius down the group, which permits a larger number of accessible vibrational configurations. Furthermore, when comparing Al-based hydrides to their Si-based analogs, the former exhibit higher S$_{vib}$, indicating a greater degree of thermal energy dispersion and intrinsic disorder. This can be attributed to the larger atomic radius of Al relative to Si. The greater atomic size of Al introduces more spatial flexibility within the hydride lattice, enabling a wider range of vibrational modes and configurations. As a result, a greater fraction of the thermal energy is statistically spread across numerous vibrational degrees of freedom, diminishing the portion of energy available to perform useful work in the Al-based hydrides. In thermodynamics and statistical mechanics, the C$_V$ is essential for characterizing lattice vibrations, particularly across two key temperature regimes relative to the Debye temperature ($\Theta_D$) \cite{Debye1912c,Schroeder1999d}.
\par At temperatures well below the $\Theta_D$ (T $<<$ $\Theta_D$), the heat capacity follows the Debye T$^3$ law, varying as:
\begin{equation}
	C_V=\frac{12}{5}\pi^4nR\bigg(\frac{T}{\Theta_D}\bigg)^3
\end{equation}
\par In the high-temperature regime (T $>>$ $\Theta_D$), the heat capacity approaches a constant value, consistent with the classical Dulong-Petit law, which states that the molar heat capacity of a solid tends toward:
\begin{equation}
	C_V=3nR
\end{equation}
here, R = 8.314 J K$^{-1}$ mol$^{-1}$ is the gas constant.
\par As shown in figures \ref{figure3} and \ref{figure4}, the C$_V$ curves for all considered hydrides follow the expected thermodynamic behavior. At low temperatures (T $<<$ $\Theta_D$), C$_V$ increases proportionally to T$^3$, consistent with Debye's T$^3$ law. At high temperatures (T $>>$ $\Theta_D$), C$_V$ approaches a constant value, aligning with the classical Dulong-Petit limit. These trends confirm that the calculations accurately capture the specific heat behavior across both regimes, in agreement with fundamental thermodynamic principles. The linear thermal expansion coefficient ($\alpha_L$), further supports this consistency. Al-based hydrides exhibit greater responsiveness to thermal expansion compared to Si-based hydrides, indicating more pronounced lattice changes under heating. Below $\Theta_D$, $\alpha_L$ rises rapidly for all the considered materials, reflecting high sensitivity of the lattice to thermal vibrations before expansion saturates. Above $\Theta_D$, $\alpha_L$ levels off, indicating that the lattice has reached a quasi-saturated state where further temperature increases minimally affect expansion.
\subsection{Application as Solid-State Hydrogen Storage Materials}
One major impediment to the widespread use of hydrogen as a renewable energy source is the limited availability of storage systems with high gravimetric capacities. Addressing this limitation necessitates the discovery and development of advanced materials capable of storing substantial amounts of hydrogen while being stable and reversible under normal operating conditions. Hydrogen may be stored as a gas, liquid, or solid; among these, solid-state storage, particularly metal hydride-based vacancy-ordered perovskites has garnered increasing attention due to its inherent safety and high volumetric efficiency. The design and optimization of such materials is guided by two key metrics: gravimetric capacity (C$_{wt}$\%), which quantifies the mass percentage of hydrogen relative to the host material, and volumetric density ($\rho_V$), which measures the quantity of hydrogen per unit volume. These parameters are essential in assessing the real-world feasibility. The C$_{wt}$\% and $\rho_V$ of the studied Al- and Si-based hydrides are estimated using the equation given below \cite{Baaddi2023b,Fatima2023b}:
\begin{equation}
		\begin{split}
			C_{wt}\%=\bigg(\frac{\frac{H}{M}m_H}{m_{Host}+\frac{H}{M}m_H}\times100\bigg)\% \\
			\rho_{V}=\frac{N_Hm_H}{V(L)N_A}\\
		\end{split}
\end{equation}
Here, $\frac{H}{M}$ is ratio of hydrogen to material atom, m$_H$ molar mass of hydrogen, m$_{Host}$ molar mass of the host material, N$_H$ number of hydrogen absorbed, V(L) volume in liter, and N$_A$ the Avogadro number. 
\begin{table}[hbt!]
	\small
	\centering
	\caption{\ The gravimetric storage density (C$_{wt}$\%), volumetric storage density ($\rho_V$) in g.H$_2$ L$^{-1}$ and the desorption temperature (T$_D$) in K for A$_2$BH$_6$ (A = Li, Na, and K; B = Al, and Si) VODPH.}
	\label{table3}\renewcommand{\arraystretch}{1.55}
	\begin{tabular*}{0.9\textwidth}{@{\extracolsep{\fill}}l|llll|llll}
		\hline
	    &&A$_2$AlH$_6$&&&&A$_2$SiH$_6$&&\\ \hline
		A = & C$_{wt}$ & $\rho_V$ & T$_D^{GGA}$ & T$_D^{HSE06}$ &  C$_{wt}$ & $\rho_V$ & T$_D^{GGA}$ & T$_D^{HSE06}$\\ \hline    
		Li& 12.89 & 142.85 &487.23 & 524.14 & 12.60 & 144.29 & 568.43 & 642.25\\
		Na& 7.65  & 109.51 &546.28 & 553.66 & 7.55  & 111.83 & 649.63 & 701.31\\
		K & 5.44  & 80.53  &671.78 & 583.19 & 5.38  & 85.61  & 804.66 & 767.75\\
		\hline
	\end{tabular*}
\end{table}
\par As summarized in table \ref{table3}, the calculated C$_{wt}$\% and $\rho_V$ reveal notable trends between Al- and Si-based hydrides. The Al-based hydrides exhibit slightly higher C$_{wt}$\% values compared to their Si-based analogues, which can be primarily attributed to the lower atomic mass of Al relative to Si. This lower atomic weight enables a higher hydrogen-to-mass ratio, thereby enhancing the gravimetric storage performance. Conversely, Si-based hydrides tend to exhibit superior $\rho_V$, owing to their more compact crystal structures and smaller unit cell volumes. Silicon typically forms stronger and shorter covalent bonds, which contribute to denser packing and, consequently, a higher hydrogen content per unit volume. These contrasting trends highlight the fundamental trade-off between gravimetric and volumetric efficiency in hydrogen storage materials, emphasizing the importance of simultaneously optimizing atomic composition and crystal structure. Furthermore, a systematic reduction in both C$_{wt}$\% and $\rho_V$ is observed as the alkali metal component (A) is varied from Li $\rightarrow$ K. This decline can be explained by the progressive increase in atomic mass and ionic radius down the group. Heavier and larger alkali metals not only dilute the hydrogen mass fraction but also expand the crystal lattice, leading to reduced packing efficiency and lower volumetric hydrogen densities. Despite this, it is noteworthy that all the investigated compounds exhibit $\rho_V$ exceeding 40 g.H$_2$ L$^{-1}$, the target set by the U.S. Department of Energy (DOE). However, none of the hydrides meet the DOE's benchmark gravimetric capacity of 5.5 \% when A = K, indicating a clear decline in storage performance with heavier alkali metals. Overall, the results suggest that Al-based hydrides possess a more favorable balance between gravimetric and volumetric storage metrics, positioning them as more promising candidates for solid-state hydrogen storage. Among all the studied materials, Li$_2$AlH$_6$ and Na$_2$SiH$_6$ stand out for their exceptional hydrogen storage performance, combining relatively high C$_{wt}$\% with $\rho_V$ values that meet or exceed DOE targets. The evaluated C$_{wt}$\% values substantially exceed those of recently reported ordered double perovskite hydrides, specifically Mg$_2$KNaH$_6$, Ca$_2$KNaH$_6$, and K$_2$NaAlH$_6$, which exhibit comparatively lower C$_{wt}$\% of 5.19\%, 4.09\%, and 4.47\%, respectively  \cite{Rehman2024,Baaddi2023b}. This pronounced enhancement underscores the superior compositional efficiency of vacancy-ordered double perovskite hydrides (A$_2$BH$_6$-type) in comparison to their ordered counterparts (A$_2$BB'H$_6$-type). Furthermore, in comparison with conventional perovskite hydrides (ABH$_3$-type), such as KMgH$_3$, KBeH$_3$, ZrZnH$_3$, ZrCdH$_3$, CaScH$_3$, and MgScH$3$, the calculated C$_{wt}$\% values surpass those of all aforementioned materials, further reinforcing the exceptional potential of the vacancy-ordered A$_2$BH$_6$-type hydrides for high-performance hydrogen storage applications \cite{Rahman2024a,Anupam2024,Masood2024}.    
\par In designing next-generation hydrogen storage materials, the desorption temperature (T$_D$) is a critical parameter, as it governs hydrogen release efficiency. An optimal T$_D$ enables hydrogen desorption under moderate, practical conditions, ensuring both efficiency and operational safety. Excessively high T$_D$ leads to energy-inefficient release, while too low a T$_D$ risks premature desorption and potential hazards. Thus, identifying materials with suitable T$_D$ is vital for advancing hydrogen-based technologies in transportation, power generation, and portable applications. The standard Gibbs energy (E$_G$), a fundamental thermodynamic parameter used to evaluate material stability and hydrogen storage efficiency, reflecting the spontaneity of hydrogenation and dehydrogenation reactions. It is defined for a reaction as:
\begin{equation}
	E_G=\Delta H-(T_D \times \Delta S)
\end{equation}
Here, $\Delta$H and $\Delta$S represent the enthalpy and entropy changes during dehydrogenation. As temperature increases, the $\Delta$S becomes more dominant. At the T$_D$, it surpasses $\Delta$H, driving the E$_G$ to approach zero. This equilibrium condition marks the onset of hydrogen release, where absorption and desorption rates are equal. Thus, at T = T$_D$:
\begin{equation}
	T_D=\frac{\Delta H}{\Delta S}
	\label{equation3}
\end{equation}
\par At constant pressure, equation \ref{equation3} delineates the conditions under which hydrogen is released from a system. At T = T$_D$, hydrogen desorption becomes thermodynamically favorable. The interplay between enthalpy and entropy plays a pivotal role in the design of hydrogen storage materials, as it dictates the temperature range over which efficient hydrogen release can occur. A key research objective is to lower T$_D$ without compromising the storage capacity or the rate of release. In solid-state systems, the $\Delta$S associated with hydrogen release is significantly lower than in gaseous systems, due to the greater degree of disorder in gases upon heating, which is key advantage of solid-state systems. In a decomposition reaction, the $\Delta$S is primarily driven by the production of hydrogen. Under standard conditions $\Delta$S of hydrogen gas is approximately 130.7 J K$^{-1}$ mol$^{-1}$, which corresponds to the $\Delta$S associated with the formation of one mole of hydrogen gas \cite{Ain2024b}. 
\par In table \ref{table3}, we present the computed T$_D$ using both GGA and HSE06 approaches for the investigated A$_2$BH$_6$ VODPH materials. The results exhibit systematic trends with A- and B-site substitutions. Notably, T$_D$ increases progressively from Li $\rightarrow$ K across both Al- and Si-based hydrides, primarily due to the increasing ionic radii and atomic masses, which expand the lattice and enhance hydrogen bonding environments. This structural stabilization raises the thermal energy required to destabilize the hydride lattice and liberate hydrogen, thereby raising the T$_D$. The Si-based hydrides consistently exhibit higher T$_D$ than their Al-based counterparts, reflecting the stronger covalent nature of Si-H bonds compared to the more metallic Al-H interactions. For instance, Li$_2$SiH$_6$ exhibits a T$_D$ of 642.25 K (HSE06), markedly above the 524.14 K recorded for Li$_2$AlH$_6$. This distinction highlights a fundamental trade-off in material design: while stronger bonds confer greater structural stability, they also elevate the energy barrier for hydrogen release. The application of the hybrid-HSE06 functional introduces nuanced shifts in the predicted T$_D$ values, generally increasing accuracy via improved treatment of electron localization and exchange interactions. For lighter alkali metals (Li, Na), HSE06 predicts higher T$_D$ compared to GGA, reaffirming the enhanced bonding and structural stability of these compounds. However, for heavier A-site cations (e.g., A = K), HSE06 results in slightly lower T$_D$, suggesting complex, composition-dependent electronic effects.
\par From a practical standpoint, a T$_D$ range of 300-600 K is optimal for balancing thermal efficiency and safety in hydrogen storage. In this context, Li$_2$AlH$_6$ and Li$_2$SiH$_6$ show promising performance, offering a good compromise between T$_D$, C$_{wt}$\%, and $\rho_V$. Also, Na$_2$AlH$_6$ and Na$_2$SiH$_6$, with a slightly higher T$_D$, offers improved thermal stability and an acceptable C$_{wt}$\% and $\rho_V$, making them suitable for stationary or high-temperature applications. In contrast, K-based hydrides, particularly K$_2$SiH$_6$ with a T$_D$ of 767.75 K, may be less practical due to its excessive T$_D$ and lower C$_{wt}$\%. Overall, the Al- and Si-based hydrides exemplify the balance, combining favorable T$_D$ values with strong storage metrics, when A = Li and Na. Thus emerge as great promise for integration into next-generation solid-state hydrogen storage systems.
\section{Conclusion}
In summary, this study delivers a rigorous and dual-faceted analysis of the hydrogen storage and thermoelectric properties of selected A$_2$BH$_6$-type double perovskite hydrides (where A = Li, Na, and K; B = Al, and Si), employing both conventional GGA and hybrid-DFT within the HSE06 functional framework. Structural optimization confirms that all investigated hydrides adopt a cubic crystal structure belonging to the $Fm$-3$m$ space group. To assess their practical viability, thermodynamic and mechanical stability were confirmed via negative formation enthalpies and compliance with the Born mechanical stability criteria. Electronic structure analysis indicates a distinct divergence in electronic behavior: Al-based hydrides exhibit metallic characteristics, while Si-based analogues demonstrate semi-conducting behavior, with substantial band gap enlargement under the hybrid-HSE06 functional. Calculation of the optical properties for the Si-based hydrides, suggested their potentiality for UV-optoelectronic devices due to the presence of the high optical absorption in the UV-region. Most of the selected hydrides satisfy the U.S. Department of Energy (DOE) targets for hydrogen storage, with the exception of the K-based compounds, K$_2$AlH$_6$ and K$_2$SiH$_6$, which deliver gravimetric storage capacities of 5.44\% and 5.38\%, respectively. Among the series, Li$_2$AlH$_6$ and Li$_2$SiH$_6$ emerge as the two most promising candidates, exhibiting superior hydrogen storage capacities and favorable hydrogen desorption temperatures. Additionally, these materials display promising thermoelectric behavior, with very low lattice thermal conductivity ($\kappa_L$ in the order of 10$^{-1}$ W m$^{-1}$ K$^{-1}$), and the dimensionless figure of merit (ZT) values ranging from 0.65 to 0.90 within the temperature range of 400-800 K, underscoring their potential for integration into advanced thermoelectric technologies

\section{Acknowledgments}
R. Zosiamliana gratefully acknowledges the assistance received from the "National Fellowship and Scholarship for Higher Education of ST Students (NFST), Ministry of Tribal Affairs, Government of India". Award No.: 202223-NFST-MIZ-01557 (dated 28.06.2023).
\par Lalhriat Zuala acknowledges support for this research work provided by Anusandhan National Research Foundation (ANRF), Government of India, vide Grant No.: ANRF/IRG/ 2024/000695/PS (dated 22.3.2025).
\par A. Laref acknowledges support from the "Research Center of the Female Scientific and Medical Colleges",  Deanship of Scientific Research, King Saud University.
\par The research is partially funded by the Ministry of Science and Higher Education of the Russian Federation as part of the World-Class Research Center program: Advanced Digital Technologies (contract No. 075-15-2022-312 dated 20.04.2022).

\bibliographystyle{elsarticle-num-names} 
\bibliography{A2BH6}

\begin{thebibliography}{59}
\expandafter\ifx\csname natexlab\endcsname\relax\def\natexlab#1{#1}\fi
\providecommand{\url}[1]{\texttt{#1}}
\providecommand{\href}[2]{#2}
\providecommand{\path}[1]{#1}
\providecommand{\DOIprefix}{doi:}
\providecommand{\ArXivprefix}{arXiv:}
\providecommand{\URLprefix}{URL: }
\providecommand{\Pubmedprefix}{pmid:}
\providecommand{\doi}[1]{\href{http://dx.doi.org/#1}{\path{#1}}}
\providecommand{\Pubmed}[1]{\href{pmid:#1}{\path{#1}}}
\providecommand{\bibinfo}[2]{#2}
\ifx\xfnm\relax \def\xfnm[#1]{\unskip,\space#1}\fi
\bibitem[{Baxter et~al.(2009)Baxter, Bian, Chen, Danielson, Dresselhaus,
  Fedorov, Fisher, Jones, Maginn, Kortshagen, Manthiram, Nozik, Rolison, Sands,
  Shi, Sholl, and Wu}]{Baxter2009}
\bibinfo{author}{J.~Baxter}, \bibinfo{author}{Z.~Bian},
  \bibinfo{author}{G.~Chen}, \bibinfo{author}{D.~Danielson},
  \bibinfo{author}{M.~S. Dresselhaus}, \bibinfo{author}{A.~G. Fedorov},
  \bibinfo{author}{T.~S. Fisher}, \bibinfo{author}{C.~W. Jones},
  \bibinfo{author}{E.~Maginn}, \bibinfo{author}{U.~Kortshagen},
  \bibinfo{author}{A.~Manthiram}, \bibinfo{author}{A.~Nozik},
  \bibinfo{author}{D.~R. Rolison}, \bibinfo{author}{T.~Sands},
  \bibinfo{author}{L.~Shi}, \bibinfo{author}{D.~Sholl},
  \bibinfo{author}{Y.~Wu},
\newblock \bibinfo{title}{{Nanoscale design to enable the revolution in
  renewable energy}},
\newblock \bibinfo{journal}{Energy \& Environmental Science}
  \bibinfo{volume}{2} (\bibinfo{year}{2009}) \bibinfo{pages}{559}.
  \DOIprefix\doi{10.1039/b821698c}.
\bibitem[{Sadeq et~al.(2024)Sadeq, Homod, Hussein, Togun, Mahmoodi, Isleem,
  Patil, and Moghaddam}]{Sadeq2024}
\bibinfo{author}{A.~M. Sadeq}, \bibinfo{author}{R.~Z. Homod},
  \bibinfo{author}{A.~K. Hussein}, \bibinfo{author}{H.~Togun},
  \bibinfo{author}{A.~Mahmoodi}, \bibinfo{author}{H.~F. Isleem},
  \bibinfo{author}{A.~R. Patil}, \bibinfo{author}{A.~H. Moghaddam},
\newblock \bibinfo{title}{{Hydrogen energy systems: Technologies, trends, and
  future prospects.}},
\newblock \bibinfo{journal}{The Science of the total environment}
  \bibinfo{volume}{939} (\bibinfo{year}{2024}) \bibinfo{pages}{173622}.
  \DOIprefix\doi{10.1016/j.scitotenv.2024.173622}.
\bibitem[{Hassan et~al.(2024)Hassan, Viktor, {J. Al-Musawi}, {Mahmood Ali},
  Algburi, Alzoubi, {Khudhair Al-Jiboory}, {Zuhair Sameen}, Salman, and
  Jaszczur}]{Hassan2024}
\bibinfo{author}{Q.~Hassan}, \bibinfo{author}{P.~Viktor},
  \bibinfo{author}{T.~{J. Al-Musawi}}, \bibinfo{author}{B.~{Mahmood Ali}},
  \bibinfo{author}{S.~Algburi}, \bibinfo{author}{H.~M. Alzoubi},
  \bibinfo{author}{A.~{Khudhair Al-Jiboory}}, \bibinfo{author}{A.~{Zuhair
  Sameen}}, \bibinfo{author}{H.~M. Salman}, \bibinfo{author}{M.~Jaszczur},
\newblock \bibinfo{title}{{The renewable energy role in the global energy
  Transformations}},
\newblock \bibinfo{journal}{Renewable Energy Focus} \bibinfo{volume}{48}
  (\bibinfo{year}{2024}) \bibinfo{pages}{100545}.
  \DOIprefix\doi{10.1016/j.ref.2024.100545}.
\bibitem[{Bertaglia et~al.(2024)Bertaglia, Costa, Lanceros-M{\'{e}}ndez, and
  Crespilho}]{Bertaglia2024}
\bibinfo{author}{T.~Bertaglia}, \bibinfo{author}{C.~M. Costa},
  \bibinfo{author}{S.~Lanceros-M{\'{e}}ndez}, \bibinfo{author}{F.~N.
  Crespilho},
\newblock \bibinfo{title}{{Eco-friendly, sustainable, and safe energy storage:
  a nature-inspired materials paradigm shift}},
\newblock \bibinfo{journal}{Materials Advances} \bibinfo{volume}{5}
  (\bibinfo{year}{2024}) \bibinfo{pages}{7534--7547}.
  \DOIprefix\doi{10.1039/D4MA00363B}.
\bibitem[{Nnabuife et~al.(2025)Nnabuife, Hamzat, Whidborne, Kuang, and
  Jenkins}]{Nnabuife2025}
\bibinfo{author}{S.~G. Nnabuife}, \bibinfo{author}{A.~K. Hamzat},
  \bibinfo{author}{J.~Whidborne}, \bibinfo{author}{B.~Kuang},
  \bibinfo{author}{K.~W. Jenkins},
\newblock \bibinfo{title}{{Integration of renewable energy sources in tandem
  with electrolysis: A technology review for green hydrogen production}},
\newblock \bibinfo{journal}{International Journal of Hydrogen Energy}
  \bibinfo{volume}{107} (\bibinfo{year}{2025}) \bibinfo{pages}{218--240}.
  \DOIprefix\doi{10.1016/j.ijhydene.2024.06.342}.
\bibitem[{Mahmood et~al.(2021)Mahmood, Hassan, Flemban, {Ul Haq}, AlFaify,
  Kattan, and Laref}]{Mahmood2021}
\bibinfo{author}{Q.~Mahmood}, \bibinfo{author}{M.~Hassan},
  \bibinfo{author}{T.~H. Flemban}, \bibinfo{author}{B.~{Ul Haq}},
  \bibinfo{author}{S.~AlFaify}, \bibinfo{author}{N.~A. Kattan},
  \bibinfo{author}{A.~Laref},
\newblock \bibinfo{title}{{Optoelectronic and thermoelectric properties of
  double perovskite Rb2PtX6 (X = Cl, Br) for energy harvesting:
  First-principles investigations}},
\newblock \bibinfo{journal}{Journal of Physics and Chemistry of Solids}
  \bibinfo{volume}{148} (\bibinfo{year}{2021}) \bibinfo{pages}{109665}.
  \DOIprefix\doi{10.1016/j.jpcs.2020.109665}.
\bibitem[{Mustafa et~al.(2024)Mustafa, Younas, Alkhaldi, Mera, Alqorashi,
  Hakami, Mahmoud, Boukhris, and Mahmood}]{Mustafa2024}
\bibinfo{author}{G.~M. Mustafa}, \bibinfo{author}{B.~Younas},
  \bibinfo{author}{H.~D. Alkhaldi}, \bibinfo{author}{A.~Mera},
  \bibinfo{author}{A.~K. Alqorashi}, \bibinfo{author}{J.~Hakami},
  \bibinfo{author}{S.~A. Mahmoud}, \bibinfo{author}{I.~Boukhris},
  \bibinfo{author}{Q.~Mahmood},
\newblock \bibinfo{title}{{First principle study of physical aspects and
  hydrogen storage capacity of magnesium-based double perovskite hydrides
  Mg2XH6 (X = Cr, Mn)}},
\newblock \bibinfo{journal}{International Journal of Hydrogen Energy}
  \bibinfo{volume}{95} (\bibinfo{year}{2024}) \bibinfo{pages}{300--308}.
  \DOIprefix\doi{10.1016/j.ijhydene.2024.11.112}.
\bibitem[{Zelai(2024)}]{Zelai2024}
\bibinfo{author}{T.~Zelai},
\newblock \bibinfo{title}{{Study of magnetic, thermoelectric, and mechanical
  properties of double perovskites Be2XH6 (X = Cr and Mn) for spintronic and
  hydrogen-storage applications}},
\newblock \bibinfo{journal}{Inorganic Chemistry Communications}
  \bibinfo{volume}{165} (\bibinfo{year}{2024}) \bibinfo{pages}{112579}.
  \DOIprefix\doi{10.1016/j.inoche.2024.112579}.
\bibitem[{Ayyaz et~al.(2025)Ayyaz, {Abaid Ullah}, Zaman, Alkhaldi, Mahmood,
  Boukhris, Al-Buriahi, and mana Al-Anazy}]{Ayyaz2025}
\bibinfo{author}{A.~Ayyaz}, \bibinfo{author}{M.~{Abaid Ullah}},
  \bibinfo{author}{M.~Zaman}, \bibinfo{author}{N.~D. Alkhaldi},
  \bibinfo{author}{Q.~Mahmood}, \bibinfo{author}{I.~Boukhris},
  \bibinfo{author}{M.~Al-Buriahi}, \bibinfo{author}{M.~mana Al-Anazy},
\newblock \bibinfo{title}{{Investigation of hydrogen storage and energy
  harvesting potential of double perovskite hydrides A2LiCuH6 (A =
  Be/Mg/Ca/Sr): A DFT approach}},
\newblock \bibinfo{journal}{International Journal of Hydrogen Energy}
  \bibinfo{volume}{102} (\bibinfo{year}{2025}) \bibinfo{pages}{1329--1339}.
  \DOIprefix\doi{10.1016/j.ijhydene.2025.01.117}.
\bibitem[{El-Shafie et~al.(2019)El-Shafie, Kambara, and
  Hayakawa}]{El-Shafie2019}
\bibinfo{author}{M.~El-Shafie}, \bibinfo{author}{S.~Kambara},
  \bibinfo{author}{Y.~Hayakawa},
\newblock \bibinfo{title}{{Hydrogen Production Technologies Overview}},
\newblock \bibinfo{journal}{Journal of Power and Energy Engineering}
  \bibinfo{volume}{07} (\bibinfo{year}{2019}) \bibinfo{pages}{107--154}.
  \DOIprefix\doi{10.4236/jpee.2019.71007}.
\bibitem[{El-Emam and {\"{O}}zcan(2019)}]{El-Emam2019}
\bibinfo{author}{R.~S. El-Emam}, \bibinfo{author}{H.~{\"{O}}zcan},
\newblock \bibinfo{title}{{Comprehensive review on the techno-economics of
  sustainable large-scale clean hydrogen production}},
\newblock \bibinfo{journal}{Journal of Cleaner Production}
  \bibinfo{volume}{220} (\bibinfo{year}{2019}) \bibinfo{pages}{593--609}.
  \DOIprefix\doi{10.1016/j.jclepro.2019.01.309}.
\bibitem[{Tremel et~al.(2015)Tremel, Wasserscheid, Baldauf, and
  Hammer}]{Tremel2015}
\bibinfo{author}{A.~Tremel}, \bibinfo{author}{P.~Wasserscheid},
  \bibinfo{author}{M.~Baldauf}, \bibinfo{author}{T.~Hammer},
\newblock \bibinfo{title}{{Techno-economic analysis for the synthesis of liquid
  and gaseous fuels based on hydrogen production via electrolysis}},
\newblock \bibinfo{journal}{International Journal of Hydrogen Energy}
  \bibinfo{volume}{40} (\bibinfo{year}{2015}) \bibinfo{pages}{11457--11464}.
  \DOIprefix\doi{10.1016/j.ijhydene.2015.01.097}.
\bibitem[{Boateng et~al.(2023)Boateng, Thiruppathi, Hung, Chow, Sridhar, and
  Chen}]{Boateng2023}
\bibinfo{author}{E.~Boateng}, \bibinfo{author}{A.~R. Thiruppathi},
  \bibinfo{author}{C.-K. Hung}, \bibinfo{author}{D.~Chow},
  \bibinfo{author}{D.~Sridhar}, \bibinfo{author}{A.~Chen},
\newblock \bibinfo{title}{{Functionalization of graphene-based nanomaterials
  for energy and hydrogen storage}},
\newblock \bibinfo{journal}{Electrochimica Acta} \bibinfo{volume}{452}
  (\bibinfo{year}{2023}) \bibinfo{pages}{142340}.
  \DOIprefix\doi{10.1016/j.electacta.2023.142340}.
\bibitem[{Chettri et~al.(2021)Chettri, Patra, Hieu, and Rai}]{Chettri2021b}
\bibinfo{author}{B.~Chettri}, \bibinfo{author}{P.~Patra},
  \bibinfo{author}{N.~N. Hieu}, \bibinfo{author}{D.~Rai},
\newblock \bibinfo{title}{{Hexagonal boron nitride ( <math altimg="si2.svg">
  <mi>h</mi> </math> -BN) nanosheet as a potential hydrogen adsorption
  material: A density functional theory (DFT) study}},
\newblock \bibinfo{journal}{Surfaces and Interfaces} \bibinfo{volume}{24}
  (\bibinfo{year}{2021}) \bibinfo{pages}{101043}.
  \DOIprefix\doi{10.1016/j.surfin.2021.101043}.
\bibitem[{Kumar et~al.(2021)Kumar, Singh, Hashmi, and Kim}]{Kumar2021}
\bibinfo{author}{P.~Kumar}, \bibinfo{author}{S.~Singh},
  \bibinfo{author}{S.~Hashmi}, \bibinfo{author}{K.-H. Kim},
\newblock \bibinfo{title}{{MXenes: Emerging 2D materials for hydrogen
  storage}},
\newblock \bibinfo{journal}{Nano Energy} \bibinfo{volume}{85}
  (\bibinfo{year}{2021}) \bibinfo{pages}{105989}.
  \DOIprefix\doi{10.1016/j.nanoen.2021.105989}.
\bibitem[{Bora et~al.(2024)Bora, Kumar, and Sinha}]{Bora2024}
\bibinfo{author}{P.~Bora}, \bibinfo{author}{S.~Kumar},
  \bibinfo{author}{D.~Sinha},
\newblock \bibinfo{title}{{2D transition metal dichalcogenides for efficient
  hydrogen generation}},
\newblock \bibinfo{journal}{Materials Today Sustainability}
  \bibinfo{volume}{27} (\bibinfo{year}{2024}) \bibinfo{pages}{100914}.
  \DOIprefix\doi{10.1016/j.mtsust.2024.100914}.
\bibitem[{Ledwaba et~al.(2023)Ledwaba, Karimzadeh, and Jen}]{Ledwaba2023}
\bibinfo{author}{K.~Ledwaba}, \bibinfo{author}{S.~Karimzadeh},
  \bibinfo{author}{T.-C. Jen},
\newblock \bibinfo{title}{{Emerging borophene two-dimensional nanomaterials for
  hydrogen storage}},
\newblock \bibinfo{journal}{Materials Today Sustainability}
  \bibinfo{volume}{22} (\bibinfo{year}{2023}) \bibinfo{pages}{100412}.
  \DOIprefix\doi{10.1016/j.mtsust.2023.100412}.
\bibitem[{Garara et~al.(2019)Garara, Benzidi, Lakhal, Louilidi, Ez-Zahraouy,
  {El Kenz}, Hamedoun, Benyoussef, Kara, and Mounkachi}]{Garara2019}
\bibinfo{author}{M.~Garara}, \bibinfo{author}{H.~Benzidi},
  \bibinfo{author}{M.~Lakhal}, \bibinfo{author}{M.~Louilidi},
  \bibinfo{author}{H.~Ez-Zahraouy}, \bibinfo{author}{A.~{El Kenz}},
  \bibinfo{author}{M.~Hamedoun}, \bibinfo{author}{A.~Benyoussef},
  \bibinfo{author}{A.~Kara}, \bibinfo{author}{O.~Mounkachi},
\newblock \bibinfo{title}{{Phosphorene: A promising candidate for H2 storage at
  room temperature}},
\newblock \bibinfo{journal}{International Journal of Hydrogen Energy}
  \bibinfo{volume}{44} (\bibinfo{year}{2019}) \bibinfo{pages}{24829--24838}.
  \DOIprefix\doi{10.1016/j.ijhydene.2019.07.194}.
\bibitem[{SAKINTUNA et~al.(2007)SAKINTUNA, LAMARIDARKRIM, and
  HIRSCHER}]{Sakintuna2007}
\bibinfo{author}{B.~SAKINTUNA}, \bibinfo{author}{F.~LAMARIDARKRIM},
  \bibinfo{author}{M.~HIRSCHER},
\newblock \bibinfo{title}{{Metal hydride materials for solid hydrogen storage:
  A review}},
\newblock \bibinfo{journal}{International Journal of Hydrogen Energy}
  \bibinfo{volume}{32} (\bibinfo{year}{2007}) \bibinfo{pages}{1121--1140}.
  \DOIprefix\doi{10.1016/j.ijhydene.2006.11.022}.
\bibitem[{Kalibek et~al.(2024)Kalibek, Ospanova, Suleimenova, Soltan, Orazbek,
  Makhmet, Rafikova, and Nuraje}]{Kalibek2024}
\bibinfo{author}{M.~R. Kalibek}, \bibinfo{author}{A.~D. Ospanova},
  \bibinfo{author}{B.~Suleimenova}, \bibinfo{author}{R.~Soltan},
  \bibinfo{author}{T.~Orazbek}, \bibinfo{author}{A.~M. Makhmet},
  \bibinfo{author}{K.~S. Rafikova}, \bibinfo{author}{N.~Nuraje},
\newblock \bibinfo{title}{{Solid-state hydrogen storage materials}},
\newblock \bibinfo{journal}{Discover Nano} \bibinfo{volume}{19}
  (\bibinfo{year}{2024}) \bibinfo{pages}{195}.
  \DOIprefix\doi{10.1186/s11671-024-04137-y}.
\bibitem[{Nemukula et~al.(2025)Nemukula, Mtshali, and
  Nemangwele}]{Nemukula2025}
\bibinfo{author}{E.~Nemukula}, \bibinfo{author}{C.~B. Mtshali},
  \bibinfo{author}{F.~Nemangwele},
\newblock \bibinfo{title}{{Metal Hydrides for Sustainable Hydrogen Storage: A
  Review}},
\newblock \bibinfo{journal}{International Journal of Energy Research}
  \bibinfo{volume}{2025} (\bibinfo{year}{2025}) \bibinfo{pages}{6300225}.
  \DOIprefix\doi{10.1155/er/6300225}.
\bibitem[{Hakami and Alathlawi(2024)}]{Hakami2024}
\bibinfo{author}{O.~Hakami}, \bibinfo{author}{H.~J. Alathlawi},
\newblock \bibinfo{title}{{Study of mechanical, optoelectronic, and
  thermoelectric characteristics of Be/Mg ions Based double perovskites A2FeH6
  (A= Be, Mg) for hydrogen storage applications}},
\newblock \bibinfo{journal}{International Journal of Hydrogen Energy}
  \bibinfo{volume}{83} (\bibinfo{year}{2024}) \bibinfo{pages}{307--316}.
  \DOIprefix\doi{10.1016/j.ijhydene.2024.08.143}.
\bibitem[{Alkhaldi(2025)}]{Alkhaldi2025}
\bibinfo{author}{N.~D. Alkhaldi},
\newblock \bibinfo{title}{{Study of hydrogen storage potential, optoelectronic
  and thermoelectric response of double perovskites hydrides X2CuAlH6 (X = Li,
  Na) for renewable energy}},
\newblock \bibinfo{journal}{Inorganic Chemistry Communications}
  \bibinfo{volume}{176} (\bibinfo{year}{2025}) \bibinfo{pages}{114208}.
  \DOIprefix\doi{10.1016/j.inoche.2025.114208}.
\bibitem[{Hafner(2008)}]{Hafner2008a}
\bibinfo{author}{J.~Hafner},
\newblock \bibinfo{title}{{Ab‐initio simulations of materials using VASP:
  Density‐functional theory and beyond}},
\newblock \bibinfo{journal}{Journal of Computational Chemistry}
  \bibinfo{volume}{29} (\bibinfo{year}{2008}) \bibinfo{pages}{2044--2078}.
  \DOIprefix\doi{10.1002/jcc.21057}.
\bibitem[{Kresse and Furthm{\"{u}}ller(1996)}]{Kresse1996e}
\bibinfo{author}{G.~Kresse}, \bibinfo{author}{J.~Furthm{\"{u}}ller},
\newblock \bibinfo{title}{{Efficient iterative schemes for ab initio
  total-energy calculations using a plane-wave basis set}},
\newblock \bibinfo{journal}{Physical Review B} \bibinfo{volume}{54}
  (\bibinfo{year}{1996}) \bibinfo{pages}{11169--11186}.
  \DOIprefix\doi{10.1103/PhysRevB.54.11169}.
\bibitem[{Perdew et~al.(1996)Perdew, Burke, and Ernzerhof}]{Perdew1996l}
\bibinfo{author}{J.~P. Perdew}, \bibinfo{author}{K.~Burke},
  \bibinfo{author}{M.~Ernzerhof},
\newblock \bibinfo{title}{{Generalized Gradient Approximation Made Simple}},
\newblock \bibinfo{journal}{Physical Review Letters} \bibinfo{volume}{77}
  (\bibinfo{year}{1996}) \bibinfo{pages}{3865--3868}.
  \DOIprefix\doi{10.1103/PhysRevLett.77.3865}.
\bibitem[{Heyd et~al.(2003)Heyd, Scuseria, and Ernzerhof}]{Heyd2003c}
\bibinfo{author}{J.~Heyd}, \bibinfo{author}{G.~E. Scuseria},
  \bibinfo{author}{M.~Ernzerhof},
\newblock \bibinfo{title}{{Hybrid functionals based on a screened Coulomb
  potential}},
\newblock \bibinfo{journal}{The Journal of Chemical Physics}
  \bibinfo{volume}{118} (\bibinfo{year}{2003}) \bibinfo{pages}{8207--8215}.
  \DOIprefix\doi{10.1063/1.1564060}.
\bibitem[{Head and Zerner(1985)}]{Head1985a}
\bibinfo{author}{J.~D. Head}, \bibinfo{author}{M.~C. Zerner},
\newblock \bibinfo{title}{{A Broyden—Fletcher—Goldfarb—Shanno
  optimization procedure for molecular geometries}},
\newblock \bibinfo{journal}{Chemical Physics Letters} \bibinfo{volume}{122}
  (\bibinfo{year}{1985}) \bibinfo{pages}{264--270}.
  \DOIprefix\doi{10.1016/0009-2614(85)80574-1}.
\bibitem[{Nawi et~al.(2006)Nawi, Ransing, and Ransing}]{Nawi2006a}
\bibinfo{author}{N.~Nawi}, \bibinfo{author}{M.~Ransing},
  \bibinfo{author}{R.~Ransing},
\newblock \bibinfo{title}{{An Improved Learning Algorithm Based on The
  Broyden-Fletcher-Goldfarb-Shanno (BFGS) Method For Back Propagation Neural
  Networks}},
\newblock in: \bibinfo{booktitle}{Sixth International Conference on Intelligent
  Systems Design and Applications}, volume~\bibinfo{volume}{1},
  \bibinfo{publisher}{IEEE}, \bibinfo{year}{2006}, pp.
  \bibinfo{pages}{152--157}. \DOIprefix\doi{10.1109/ISDA.2006.95}.
\bibitem[{Monkhorst and Pack(1976)}]{Monkhorst1976k}
\bibinfo{author}{H.~J. Monkhorst}, \bibinfo{author}{J.~D. Pack},
\newblock \bibinfo{title}{{Special points for Brillouin-zone integrations}},
\newblock \bibinfo{journal}{Physical Review B} \bibinfo{volume}{13}
  (\bibinfo{year}{1976}) \bibinfo{pages}{5188--5192}.
  \DOIprefix\doi{10.1103/PhysRevB.13.5188}.
\bibitem[{Bl{\"{o}}chl et~al.(1994)Bl{\"{o}}chl, Jepsen, and
  Andersen}]{Blochl1994d}
\bibinfo{author}{P.~E. Bl{\"{o}}chl}, \bibinfo{author}{O.~Jepsen},
  \bibinfo{author}{O.~K. Andersen},
\newblock \bibinfo{title}{{Improved tetrahedron method for Brillouin-zone
  integrations}},
\newblock \bibinfo{journal}{Physical Review B} \bibinfo{volume}{49}
  (\bibinfo{year}{1994}) \bibinfo{pages}{16223--16233}.
  \DOIprefix\doi{10.1103/PhysRevB.49.16223}.
\bibitem[{Talebi et~al.(2023)Talebi, Mokhtari, and Soleimanian}]{Talebi2023}
\bibinfo{author}{M.~Talebi}, \bibinfo{author}{A.~Mokhtari},
  \bibinfo{author}{V.~Soleimanian},
\newblock \bibinfo{title}{{Ab-inito simulation of the structural, electronic
  and optical properties for the vacancy-ordered double perovskites ATiI (A =
  Cs or NH); a time-dependent density functional theory study}},
\newblock \bibinfo{journal}{Journal of Physics and Chemistry of Solids}
  \bibinfo{volume}{176} (\bibinfo{year}{2023}) \bibinfo{pages}{111262}.
  \DOIprefix\doi{10.1016/j.jpcs.2023.111262}.
\bibitem[{Alburaih et~al.(2024)Alburaih, Nazir, Noor, Laref, and {Saad
  H.-E.}}]{Alburaih2024}
\bibinfo{author}{H.~A. Alburaih}, \bibinfo{author}{S.~Nazir},
  \bibinfo{author}{N.~A. Noor}, \bibinfo{author}{A.~Laref},
  \bibinfo{author}{M.~M. {Saad H.-E.}},
\newblock \bibinfo{title}{{Physical properties of vacancy-ordered double
  perovskites K 2 TcZ 6 (Z = Cl, Br) for spintronics applications: DFT
  calculations}},
\newblock \bibinfo{journal}{RSC Advances} \bibinfo{volume}{14}
  (\bibinfo{year}{2024}) \bibinfo{pages}{1822--1832}.
  \DOIprefix\doi{10.1039/D3RA07603B}.
\bibitem[{Hayat and Khalil(2023)}]{Hayat2023}
\bibinfo{author}{M.~S. Hayat}, \bibinfo{author}{R.~M. Khalil},
\newblock \bibinfo{title}{{A DFT engineering of double halide type perovskites
  Cs2SiCl6, Cs2GeCl6, Cs2SnCl6 for optoelectronic applications}},
\newblock \bibinfo{journal}{Solid State Communications} \bibinfo{volume}{361}
  (\bibinfo{year}{2023}) \bibinfo{pages}{115064}.
  \DOIprefix\doi{10.1016/j.ssc.2023.115064}.
\bibitem[{Renthlei et~al.(2023)Renthlei, Celestine, Kima, Zuala, Mawia,
  Chettri, Singh, Abdullaev, Ezzeldien, and Rai}]{Renthlei2023h}
\bibinfo{author}{Z.~Renthlei}, \bibinfo{author}{L.~Celestine},
  \bibinfo{author}{L.~Kima}, \bibinfo{author}{L.~Zuala},
  \bibinfo{author}{Z.~Mawia}, \bibinfo{author}{B.~Chettri},
  \bibinfo{author}{Y.~T. Singh}, \bibinfo{author}{S.~Abdullaev},
  \bibinfo{author}{M.~Ezzeldien}, \bibinfo{author}{D.~P. Rai},
\newblock \bibinfo{title}{{Theoretical Investigation of Lead Perovskite PbXO 3
  (X = Ti, Zr, and Hf) for Potential Thermoelectric Applications: Hybrid-DFT
  Approach}},
\newblock \bibinfo{journal}{Energy \& Fuels} \bibinfo{volume}{37}
  (\bibinfo{year}{2023}) \bibinfo{pages}{19831--19844}.
  \DOIprefix\doi{10.1021/acs.energyfuels.3c02797}.
\bibitem[{Zosiamliana et~al.(2025)Zosiamliana, Lalroliana, Celestine, Hima,
  Chanu, Zuala, Yvaz, and Rai}]{Zosiamliana2025a}
\bibinfo{author}{R.~Zosiamliana}, \bibinfo{author}{B.~Lalroliana},
  \bibinfo{author}{L.~Celestine}, \bibinfo{author}{L.~Hima},
  \bibinfo{author}{S.~T. Chanu}, \bibinfo{author}{L.~Zuala},
  \bibinfo{author}{A.~Yvaz}, \bibinfo{author}{D.~Rai},
\newblock \bibinfo{title}{{A systematic investigation of Li- and Na-based
  perovskite hydrides as potential hydrogen storage materials}},
\newblock \bibinfo{journal}{International Journal of Hydrogen Energy}
  \bibinfo{volume}{105} (\bibinfo{year}{2025}) \bibinfo{pages}{748--758}.
  \DOIprefix\doi{10.1016/j.ijhydene.2025.01.186}.
\bibitem[{Mulliken(1955)}]{Mulliken1955b}
\bibinfo{author}{R.~S. Mulliken},
\newblock \bibinfo{title}{{Electronic Population Analysis on LCAO–MO
  Molecular Wave Functions. I}},
\newblock \bibinfo{journal}{The Journal of Chemical Physics}
  \bibinfo{volume}{23} (\bibinfo{year}{1955}) \bibinfo{pages}{1833--1840}.
  \DOIprefix\doi{10.1063/1.1740588}.
\bibitem[{Ambrosch-Draxl and Sofo(2006)}]{Ambrosch2006}
\bibinfo{author}{C.~Ambrosch-Draxl}, \bibinfo{author}{J.~O. Sofo},
\newblock \bibinfo{title}{Linear optical properties of solids within the
  full-potential linearized augmented planewave method},
\newblock \bibinfo{journal}{Computer Physics Communications}
  \bibinfo{volume}{175} (\bibinfo{year}{2006}) \bibinfo{pages}{1--14}.
  \DOIprefix\doi{https://doi.org/10.1016/j.cpc.2006.03.005}.
\bibitem[{Wang(1996)}]{WANG1996}
\bibinfo{author}{Z.~Wang},
\newblock \bibinfo{title}{Valence electron excitations and plasmon oscillations
  in thin films, surfaces, interfaces and small particles},
\newblock \bibinfo{journal}{Micron} \bibinfo{volume}{27} (\bibinfo{year}{1996})
  \bibinfo{pages}{265--299}.
  \DOIprefix\doi{https://doi.org/10.1016/0968-4328(96)00011-X}.
\bibitem[{Mouhat and Coudert(2014)}]{Mouhat2014j}
\bibinfo{author}{F.~Mouhat}, \bibinfo{author}{F.-X. Coudert},
\newblock \bibinfo{title}{{Necessary and sufficient elastic stability
  conditions in various crystal systems}},
\newblock \bibinfo{journal}{Physical Review B} \bibinfo{volume}{90}
  (\bibinfo{year}{2014}) \bibinfo{pages}{224104}.
  \DOIprefix\doi{10.1103/PhysRevB.90.224104}.
\bibitem[{Born(1940)}]{Born1940i}
\bibinfo{author}{M.~Born},
\newblock \bibinfo{title}{{On the stability of crystal lattices. I}},
\newblock \bibinfo{journal}{Mathematical Proceedings of the Cambridge
  Philosophical Society} \bibinfo{volume}{36} (\bibinfo{year}{1940})
  \bibinfo{pages}{160--172}. \DOIprefix\doi{10.1017/S0305004100017138}.
\bibitem[{Rehman et~al.(2024)Rehman, Rehman, Usman, Alomar, Khan, and
  Fatima}]{Rehman2024}
\bibinfo{author}{M.~A. Rehman}, \bibinfo{author}{Z.~U. Rehman},
  \bibinfo{author}{M.~Usman}, \bibinfo{author}{S.~Y. Alomar},
  \bibinfo{author}{M.~J. Khan}, \bibinfo{author}{J.~Fatima},
\newblock \bibinfo{title}{{Exploring the hydrogen storage in novel perovskite
  hydrides: A DFT study}},
\newblock \bibinfo{journal}{International Journal of Hydrogen Energy}
  \bibinfo{volume}{84} (\bibinfo{year}{2024}) \bibinfo{pages}{447--456}.
  \DOIprefix\doi{10.1016/j.ijhydene.2024.08.246}.
\bibitem[{Kleinman(1962)}]{Kleinman1962h}
\bibinfo{author}{L.~Kleinman},
\newblock \bibinfo{title}{{Deformation Potentials in Silicon. I. Uniaxial
  Strain}},
\newblock \bibinfo{journal}{Physical Review} \bibinfo{volume}{128}
  (\bibinfo{year}{1962}) \bibinfo{pages}{2614--2621}.
  \DOIprefix\doi{10.1103/PhysRev.128.2614}.
\bibitem[{Chung and Buessem(1968)}]{Chung1968f}
\bibinfo{author}{D.~H. Chung}, \bibinfo{author}{W.~R. Buessem},
\newblock \bibinfo{title}{{The Voigt-Reuss-Hill (VRH) Approximation and the
  Elastic Moduli of Polycrystalline ZnO, TiO2 (Rutile), and $\alpha$-Al2O3}},
\newblock \bibinfo{journal}{Journal of Applied Physics} \bibinfo{volume}{39}
  (\bibinfo{year}{1968}) \bibinfo{pages}{2777--2782}.
  \DOIprefix\doi{10.1063/1.1656672}.
\bibitem[{Sun et~al.(2005)Sun, Music, Ahuja, and Schneider}]{Sun2005c}
\bibinfo{author}{Z.~Sun}, \bibinfo{author}{D.~Music},
  \bibinfo{author}{R.~Ahuja}, \bibinfo{author}{J.~M. Schneider},
\newblock \bibinfo{title}{{Theoretical investigation of the bonding and elastic
  properties of nanolayered ternary nitrides}},
\newblock \bibinfo{journal}{Physical Review B} \bibinfo{volume}{71}
  (\bibinfo{year}{2005}) \bibinfo{pages}{193402}.
  \DOIprefix\doi{10.1103/PhysRevB.71.193402}.
\bibitem[{Ahmed et~al.(2023)Ahmed, Mahamudujjaman, Afzal, Islam, Islam, and
  Naqib}]{Ahmed2023c}
\bibinfo{author}{R.~Ahmed}, \bibinfo{author}{M.~Mahamudujjaman},
  \bibinfo{author}{M.~A. Afzal}, \bibinfo{author}{M.~S. Islam},
  \bibinfo{author}{R.~Islam}, \bibinfo{author}{S.~Naqib},
\newblock \bibinfo{title}{{DFT based comparative analysis of the physical
  properties of some binary transition metal carbides XC (X = Nb, Ta, Ti)}},
\newblock \bibinfo{journal}{Journal of Materials Research and Technology}
  \bibinfo{volume}{24} (\bibinfo{year}{2023}) \bibinfo{pages}{4808--4832}.
  \DOIprefix\doi{10.1016/j.jmrt.2023.04.147}.
\bibitem[{Cahill and Pohl(1989)}]{Cahill1989i}
\bibinfo{author}{D.~G. Cahill}, \bibinfo{author}{R.~Pohl},
\newblock \bibinfo{title}{{Heat flow and lattice vibrations in glasses}},
\newblock \bibinfo{journal}{Solid State Communications} \bibinfo{volume}{70}
  (\bibinfo{year}{1989}) \bibinfo{pages}{927--930}.
  \DOIprefix\doi{10.1016/0038-1098(89)90630-3}.
\bibitem[{Tani et~al.(2010)Tani, Takahashi, and Kido}]{Tani2010i}
\bibinfo{author}{J.-i. Tani}, \bibinfo{author}{M.~Takahashi},
  \bibinfo{author}{H.~Kido},
\newblock \bibinfo{title}{{Lattice dynamics and elastic properties of Mg3As2
  and Mg3Sb2 compounds from first-principles calculations}},
\newblock \bibinfo{journal}{Physica B: Condensed Matter} \bibinfo{volume}{405}
  (\bibinfo{year}{2010}) \bibinfo{pages}{4219--4225}.
  \DOIprefix\doi{10.1016/j.physb.2010.07.014}.
\bibitem[{Chen and Yang(2011)}]{Chen2011l}
\bibinfo{author}{H.~Chen}, \bibinfo{author}{L.~Yang},
\newblock \bibinfo{title}{{Pressure effect on the structural and elastic
  property of Hf2InC}},
\newblock \bibinfo{journal}{Physica B: Condensed Matter} \bibinfo{volume}{406}
  (\bibinfo{year}{2011}) \bibinfo{pages}{4489--4493}.
  \DOIprefix\doi{10.1016/j.physb.2011.09.013}.
\bibitem[{Chen and Sundman(2001)}]{Chen2001c}
\bibinfo{author}{Q.~Chen}, \bibinfo{author}{B.~Sundman},
\newblock \bibinfo{title}{{Calculation of debye temperature for crystalline
  structures—a case study on Ti, Zr, and Hf}},
\newblock \bibinfo{journal}{Acta Materialia} \bibinfo{volume}{49}
  (\bibinfo{year}{2001}) \bibinfo{pages}{947--961}.
  \DOIprefix\doi{10.1016/S1359-6454(01)00002-7}.
\bibitem[{Blanco et~al.(2004)Blanco, Francisco, and Lua{\~{n}}a}]{Blanco2004g}
\bibinfo{author}{M.~Blanco}, \bibinfo{author}{E.~Francisco},
  \bibinfo{author}{V.~Lua{\~{n}}a},
\newblock \bibinfo{title}{{GIBBS: isothermal-isobaric thermodynamics of solids
  from energy curves using a quasi-harmonic Debye model}},
\newblock \bibinfo{journal}{Computer Physics Communications}
  \bibinfo{volume}{158} (\bibinfo{year}{2004}) \bibinfo{pages}{57--72}.
  \DOIprefix\doi{10.1016/j.comphy.2003.12.001}.
\bibitem[{Debye(1912)}]{Debye1912c}
\bibinfo{author}{P.~Debye},
\newblock \bibinfo{title}{{Zur Theorie der spezifischen W{\"{a}}rmen}},
\newblock \bibinfo{journal}{Annalen der Physik} \bibinfo{volume}{344}
  (\bibinfo{year}{1912}) \bibinfo{pages}{789--839}.
  \DOIprefix\doi{10.1002/andp.19123441404}.
\bibitem[{Schroeder and Pribram(1999)}]{Schroeder1999d}
\bibinfo{author}{D.~V. Schroeder}, \bibinfo{author}{J.~K. Pribram},
\newblock \bibinfo{title}{{An Introduction to Thermal Physics}},
\newblock \bibinfo{journal}{American Journal of Physics} \bibinfo{volume}{67}
  (\bibinfo{year}{1999}) \bibinfo{pages}{1284--1285}.
  \DOIprefix\doi{10.1119/1.19116}.
\bibitem[{Baaddi et~al.(2023)Baaddi, Chami, Baalla, Quaoubi, Saadi, Omari, and
  Chafi}]{Baaddi2023b}
\bibinfo{author}{M.~Baaddi}, \bibinfo{author}{R.~Chami},
  \bibinfo{author}{O.~Baalla}, \bibinfo{author}{S.~E. Quaoubi},
  \bibinfo{author}{A.~Saadi}, \bibinfo{author}{L.~E.~H. Omari},
  \bibinfo{author}{M.~Chafi},
\newblock \bibinfo{title}{{The effect of strain on hydrogen storage
  characteristics in K2NaAlH6 double perovskite hydride through first principle
  method}},
\newblock \bibinfo{journal}{Environmental Science and Pollution Research}
  \bibinfo{volume}{31} (\bibinfo{year}{2023}) \bibinfo{pages}{62056--62064}.
  \DOIprefix\doi{10.1007/s11356-023-27529-6}.
\bibitem[{Fatima et~al.(2023)Fatima, Rizwan, {Naeem Ullah}, Ali, Naeem, and
  Usman}]{Fatima2023b}
\bibinfo{author}{S.~Fatima}, \bibinfo{author}{M.~Rizwan},
  \bibinfo{author}{H.~M. {Naeem Ullah}}, \bibinfo{author}{S.~S. Ali},
  \bibinfo{author}{H.~Naeem}, \bibinfo{author}{Z.~Usman},
\newblock \bibinfo{title}{{Efficient hydrogen storage in KCaF3 using GGA and
  HSE approach}},
\newblock \bibinfo{journal}{International Journal of Hydrogen Energy}
  \bibinfo{volume}{48} (\bibinfo{year}{2023}) \bibinfo{pages}{3566--3582}.
  \DOIprefix\doi{10.1016/j.ijhydene.2022.10.187}.
\bibitem[{Rahman et~al.(2024)Rahman, Islam, Rayhan, Kabir, Alim, Uddin,
  Albaqami, Mohammad, Haldhar, and Hossain}]{Rahman2024a}
\bibinfo{author}{M.~A. Rahman}, \bibinfo{author}{S.~S. Islam},
  \bibinfo{author}{M.~A. Rayhan}, \bibinfo{author}{A.~Kabir},
  \bibinfo{author}{M.~A. Alim}, \bibinfo{author}{J.~Uddin},
  \bibinfo{author}{M.~D. Albaqami}, \bibinfo{author}{S.~Mohammad},
  \bibinfo{author}{R.~Haldhar}, \bibinfo{author}{M.~K. Hossain},
\newblock \bibinfo{title}{{Comparative analysis of KXH3(X= Mg, be) hydride
  cubic perovskites for hydrogen storage properties: A computational
  approach}},
\newblock \bibinfo{journal}{International Journal of Hydrogen Energy}
  \bibinfo{volume}{80} (\bibinfo{year}{2024}) \bibinfo{pages}{725--732}.
  \DOIprefix\doi{10.1016/j.ijhydene.2024.07.064}.
\bibitem[{Anupam et~al.(2024)Anupam, Gupta, Kumar, Panwar, and
  Diwaker}]{Anupam2024}
\bibinfo{author}{Anupam}, \bibinfo{author}{S.~L. Gupta},
  \bibinfo{author}{S.~Kumar}, \bibinfo{author}{S.~Panwar},
  \bibinfo{author}{Diwaker},
\newblock \bibinfo{title}{{Ab initio studies of newly proposed zirconium based
  novel combinations of hydride perovskites ZrXH3 (X = Zn, Cd) as hydrogen
  storage applications}},
\newblock \bibinfo{journal}{International Journal of Hydrogen Energy}
  \bibinfo{volume}{55} (\bibinfo{year}{2024}) \bibinfo{pages}{1465--1475}.
  \DOIprefix\doi{10.1016/j.ijhydene.2023.11.286}.
\bibitem[{Masood et~al.(2024)Masood, Khan, Bibi, Kanwal, Bibi, Noor, Alothman,
  Rehman, and Shafiee}]{Masood2024}
\bibinfo{author}{M.~K. Masood}, \bibinfo{author}{W.~Khan},
  \bibinfo{author}{S.~Bibi}, \bibinfo{author}{A.~Kanwal},
  \bibinfo{author}{S.~Bibi}, \bibinfo{author}{G.~Noor}, \bibinfo{author}{A.~A.
  Alothman}, \bibinfo{author}{J.~Rehman}, \bibinfo{author}{S.~A. Shafiee},
\newblock \bibinfo{title}{{Physical properties of the XScH3 (X: Ca, and Mg)
  perovskite hydrides and their hydrogen storage applications}},
\newblock \bibinfo{journal}{Journal of Physics and Chemistry of Solids}
  \bibinfo{volume}{192} (\bibinfo{year}{2024}) \bibinfo{pages}{112098}.
  \DOIprefix\doi{10.1016/j.jpcs.2024.112098}.
\bibitem[{Ain et~al.(2024)Ain, Naeem, Ali, Munir, Bibi, Ghaithan, Ahmed, and
  Qaid}]{Ain2024b}
\bibinfo{author}{Q.~Ain}, \bibinfo{author}{H.~T. Naeem},
  \bibinfo{author}{M.~Ali}, \bibinfo{author}{J.~Munir},
  \bibinfo{author}{Z.~Bibi}, \bibinfo{author}{H.~M. Ghaithan},
  \bibinfo{author}{A.~A.~A. Ahmed}, \bibinfo{author}{S.~M. Qaid},
\newblock \bibinfo{title}{{A precise prediction of structure stability and
  hydrogen storage capability of KCdH3 perovskite hydride using density
  functional theory calculations}},
\newblock \bibinfo{journal}{Journal of Energy Storage} \bibinfo{volume}{100}
  (\bibinfo{year}{2024}) \bibinfo{pages}{113734}.
  \DOIprefix\doi{10.1016/j.est.2024.113734}.

\end{thebibliography}




\end{document}